\documentclass[12pt]{article}

\usepackage[nosort]{cite}
\usepackage{epsfig}
\usepackage{amsmath}
\usepackage{amsfonts}
\usepackage{amssymb}
\usepackage{multirow}
\usepackage{array}
\usepackage{subfigure}

\newcommand{\dslash}{\not\!\partial}

\newcommand{\eps}{\varepsilon}
\newcommand{\vphi}{\varphi}

 \def    \ptj           {\mbox{$p_{T j}$}}
 \def    \etaj           {\mbox{$\eta_{ j}$}}
 \def    \ptl           {\mbox{$p_{T l}$}}
 \def    \etal           {\mbox{$\eta_{ l}$}}
 \def    \gev            {\mbox{$\mathrm{GeV}$}}

\newcommand{\etmiss}{ \not \hskip -4pt E_T}

\begin{document}

\setcounter{page}{0}
\thispagestyle{empty}



\vspace*{0.8cm}

\begin{center}
{\bf \Large {
Discovering the composite Higgs through the decay of a heavy fermion
}}
\end{center}

\vskip 20pt

\begin{center}
{\large Natascia Vignaroli}
\end{center}

\vskip 20pt

\begin{center}

{\it Department of Physics and Astronomy, Iowa State University, Ames, IA 50011, USA}
\end{center}

\vskip 50pt

\begin{abstract}
\vskip 3pt
\noindent
A possible composite nature of the Higgs could be revealed at the early stage of the LHC, by analyzing the channels where the Higgs is produced from the decay of a heavy fermion. 
The Higgs production from a singly-produced heavy bottom, in particular, proves to be a promising channel. For a value $\lambda=3$ of the Higgs coupling to a heavy bottom, for example, we find that, considering a $125$ GeV Higgs which decays into a pair of $b$-quarks, a discovery is possible at the 8 TeV LHC with $30$ fb$^{-1}$ 
if the heavy bottom is lighter than roughly $530$ GeV (while an observation is possible for heavy bottom masses up to $\simeq 650$ GeV). 
Such a relatively light heavy bottom is realistic in composite Higgs models of the type considered and, up to now, experimentally allowed.
At $\sqrt{s}=14$ TeV the LHC sensitivity on the channel increases significantly. With $\lambda=3$ a discovery can occur, with $100$ fb$^{-1}$, for heavy bottom masses up to $\simeq 1040$ GeV. In the case the heavy bottom was as light as $\simeq 500$ GeV, the 14 TeV LHC would be sensitive to the measure of the $\lambda$ coupling in basically the full range $\lambda>1$ predicted by the theory. 
\end{abstract}

\vskip 13pt
\newpage


\tableofcontents

\section{Introduction}
\label{sec:introduction}

In this paper we consider the possibility that the LHC excess around $125$ GeV \cite{lhc-higgs} could be attributed to a Higgs from a physics beyond the Standard Model. In particular we will refer to the compelling scenario where the Higgs is a bound state of a strongly interacting dynamics at the 
TeV scale \cite{Georgi_Kaplan}. It could be as light as $125$ GeV, for example, if it is also the pseudo-Goldstone boson of a spontaneously-broken global symmetry of the strong sector \cite{Contino:2003ve, Agashe:2004rs}.\\
Such scenario generally predicts strong coupling between the Higgs and the new heavy fermions emerging from the strong sector. The contribution to the composite Higgs production given by the decays of the heavy fermions is thus sizable. 
Analyses focused on this type of channels can represent a very effective test of the composite nature of the Higgs, which could be revealed already at the early stage of the LHC. The composite nature of the Higgs boson can be properly analyzed from the strong scattering of Higgs and longitudinal gauge bosons. This requires, however, a much larger amount of integrated luminosity to be collected at the LHC, about $300$ fb$^{-1}$ with $\sqrt{s}=14$ TeV \cite{contino-rattazzi}. Present searches for the Higgs boson in the ``standard" channels have begun to constraint small portions of the parameter space of composite Higgs models \cite{higgs-bound}; they are not yet sensitive, however, to decay channels as $h\to b\bar{b}$. ``Alternative" channels, as the one we are proceeding to analyze, could improve the LHC sensitivity to this decay and extend the reach on the parameter space. \\
The composite Higgs production through the decays of heavy quarks has been recently analyzed in \cite{lhcNPG}.
Two types of productions for the heavy fermions have been considered in this report: the double production of vector-like quarks (see also \cite{Azatov2012}) and the production of a heavy fermion in association with its SM partner from the decay of a heavy gluon \cite{Bini, Santiago, Kong, vignaroli_tesi}.
In this work we will consider the single production of a heavy fermion, which is mediated by the exchange 
of longitudinal electro-weak bosons \cite{Mrazek:2009yu, Servant}. This type of production has been analyzed in \cite{Mrazek:2009yu} to study the prospects of discovering the heavy fermions, without considering the Higgs production. In this paper we present a first analysis of the Higgs production mediated by the single production of a heavy fermion. As a first concrete case, we will focus our analysis on the single production of a heavy bottom ($\tilde{B}$). The bottom partner can be realistically lighter than other heavy resonances and its exchange leads to a particularly promising final state with a distinctive topology. Other interesting channels, as we will highlight in Section~\ref{sec:conclusions}, involve the single production of heavy top partners (a thorough analysis of these channels has been recently performed in \cite{vignaroli_preparation}). \footnote{Such analysis could give interesting informations on the theory, especially once results were compared with those from the $\tilde{B}$ channel (see the discussion in Sec. \ref{sec:conclusions} ).} \\
The analysis we will perform considering the single production, as first pointed out in \cite{Mrazek:2009yu}, offers, contrary to the other heavy-fermion-mediated production mechanisms, the 
possibility of measuring the Higgs coupling to heavy fermions, from which it could be possible to extract important informations on the theory, such as the value of the Yukawa coupling among composite states. A further advantage of this type of analysis resides in the fact that the signal cross section drops slowerly than that of double production with the increasing of the heavy fermion mass. Therefore, an analysis of single production is expected to be more powerful than an analysis of double production, in the case of heavy fermions with mass above roughly $600$ GeV (see also \cite{Servant}). The heavy fermion production from a heavy gluon proves to be very powerful, with the possibility for a discovery at the early LHC run \cite{Santiago2012}. It has the disadvantage, however, to require 
the existence of a heavy gluon whose mass has to be comprised in a range $[\sim m^{*}, \sim 2 m^{*}]$, where $m^{*}$ denotes the mass of the heavy fermion, and to depend on several other parameters of the model, such as the top degree of compositeness \cite{Bini, vignaroli_tesi}. The analysis we will perform, on the contrary, depends on just two parameters (in addition to the Higgs mass and branching ratios), the heavy-bottom mass and its coupling to the Higgs and electro-weak gauge bosons; it could be thus also easily generalized to other scenarios.  \\
Top partners which can be singly-produced by electroweak interactions also appear, for example, in Little Higgs models \cite{Little-Higgs}. Analyses of single production in the context of Little Higgs theories can be found in \cite{Little-Higgs-analyses}.\\

The paper is organized as follows. In Section~\ref{sec:model} we review the effective two-site model that we adopt
to study the phenomenology and we define the lagrangian which is relevant for our analysis. We derive the heavy bottom production cross section and
the relevant decay rates. In Section~\ref{sec:analysis} we study the prospects of observing $\tilde{B}\to hb$ events at the LHC.
We perform a Montecarlo simulation of the signal and the main SM backgrounds and outline a strategy to maximize 
the discovery significance. We discuss the results obtained and draw our conclusions in Section~\ref{sec:conclusions}.
Finally, more details on the effective model and the expressions for the Higgs coupling to the different heavy fermions can be found in Appendix~\ref{TS10A}.

\section{A two-site effective theory for composite Higgs models}
\label{sec:model}

At energies below the compositeness scale, assumed to be of several TeV, composite operators excite from the vacuum a tower of infinite resonances; we will work in the framework of an effective theory that includes only the lowest-lying resonances and reproduces the low-energy limit of a large set of composite Higgs models and warped extra-dimensional theories with a custodial symmetry in the bulk~\cite{Agashe:2003zs}. 
Specifically, we will adopt a ``two-site'' (TS) description \cite{Contino:2006nn}, where two sectors, the weakly-coupled sector of the elementary fields
and the composite sector, that comprises the Higgs, are linearly coupled each other through mass mixing terms \cite{Kaplan:1991dc}. This leads to a scenario of partial compositeness of the SM; after diagonalization the 
elementary/composite basis rotates to the mass eigenstate one, made up of SM and heavy states that are admixture of elementary and composite modes. 
Heavier particles have larger degrees of compositeness: heavy SM particles, like the top, are more composite while the light ones are almost elementary.


\subsection{The model}

We adopt the same two-site TS10 model introduced in ref. \cite{vignaroli_tesi}. More details on the model can be found in \cite{vignaroli_bsGamma}.
The two building blocks of the model are the elementary sector and the composite sector. The composite sector has
a $SU(3)_c \times SO(4) \times U(1)_X$ global symmetry, with $SO(4) \sim SU(2)_L \times SU(2)_R$. The particle content of the elementary sector
is that of the SM without the Higgs, and the $SU(3)_c \times SU(2)_L \times U(1)_Y$ elementary fields gauge the corresponding global
invariance of the strong dynamics, with $Y = T_{R}^3 + X$. 
The composite sector comprises   
the composite Higgs
\begin{equation} \label{eq:higgs}
{\cal H} = (\mathbf{1},\mathbf{2},\mathbf{2})_{0} = 
 \begin{bmatrix} \phi_0^\dagger & \phi^+ \\ - \phi^- & \phi_0 \end{bmatrix} \, ,
\end{equation}
and the following set of vector-like composite fermions:

\begin{align}\label{eq:fermions} 
\begin{split}	
&	\mathcal{Q}_{2/3}=\left[\begin{array}{cc}
	T & T_{5/3} \\ 
	B & T_{2/3} \end{array}\right]=\left(3,2,2\right)_{2/3} \\[0.3 cm] 	
& \mathcal{\tilde{Q}}_{2/3}=\left(\begin{array}{c}
	\tilde{T}_{5/3} \\ 
	\tilde{T}\\
        \tilde{B} \end{array}\right)=\left(3,1,3\right)_{2/3}\ , \  \mathcal{\tilde{Q}'}_{2/3}=\left(\begin{array}{c}
	\tilde{T}'_{5/3} \\ 
	\tilde{T}' \\
        \tilde{B}' \end{array}\right)=\left(3,3,1\right)_{2/3}  
        \end{split}
\end{align}

The quantum numbers of the composite fermions and the Higgs under $SU(3)_c\times SU(2)_L \times SU(2)_R \times U(1)_X$ are those
specified in  eqs.(\ref{eq:higgs}), (\ref{eq:fermions}).

The fermions, in particular, can be arranged in a 10 of $SO(5)$, in fact this theory can describe the low-energy 
regime of the minimal composite Higgs model MCHM10 introduced in Ref.~\cite{Contino:2006qr}. \\

The lagrangian that describes our effective theory
is the following (we work in the gauge-less limit, and omit the terms involving the $SU(3)_c\times SU(2)_L \times U(1)_Y$ elementary gauge fields, which play no role in our analysis. We also neglect the terms involving fermions of the 1st and 2nd generations):
%
\begin{align}  \label{eq:Ltotal}
{\cal L} =&  \, {\cal L}_{elementary} + {\cal L}_{composite} + {\cal L}_{mixing} \\[0.7cm]
\label{eq:Lelem}
 {\cal L}_{elementary} =& \,\bar q_L i\!\dslash\, q_L + \bar t_R i\!\dslash\, t_R +  \bar b_R i\!\dslash\, b_R  \\[0.5cm]
\label{eq:Lcomp}
\begin{split} 
 {\cal L}_{composite}  =&  
  +\text{Tr}\left\{\bar{\mathcal{Q}}\left(i\dslash -M_{Q*}\right)\mathcal{Q}\right\}+\text{Tr}\left\{\mathcal{\bar{\tilde{Q}}}\left(i\dslash-M_{\tilde{Q}*}\right)\mathcal{\tilde{Q}}\right\}
+\text{Tr}\left\{\mathcal{\bar{\tilde{Q}}'}\left(i\dslash-M_{\tilde{Q}*}\right)\mathcal{\tilde{Q}}'\right\}  \\[0.2cm]
  & + \frac{1}{2} \,\text{Tr}\left\{  \partial_\mu {\cal H}^\dagger \partial^\mu {\cal H} \right\} - V( {\cal H}^\dagger {\cal H})
      +Y_{*}\text{Tr}\left\{\mathcal{H}\bar{\mathcal{Q}}\mathcal{\tilde{Q}'}\right\}+Y_{*}\text{Tr}\left\{\bar{\mathcal{Q}}\mathcal{H}\mathcal{\tilde{Q}}\right\}
\end{split} \\[0.5cm]
\begin{split} \label{eq:Lmixing}
{\cal L}_{mixing} =&  -\Delta_{L}\bar{q}_{L}\left(T,B\right)-\Delta_{R1}\bar{t}_{R}\tilde{T}-\Delta_{R2}\bar{b}_{R}\tilde{B}+h.c.\ .
\end{split}
\end{align}
%
where $V( {\cal H}^\dagger {\cal H})$ is the Higgs potential. 
$Y_{*}$ denotes the Yukawa coupling among composite states; as all the couplings among composites, it is assumed to be strong, $1< Y_{*} \ll 4\pi$, where $4\pi$ marks out the non-perturbative regime.
By construction, the elementary fields couple to the composite ones only through the mass mixing lagrangian ${\cal L}_{mixing}$.
This implies that the SM Yukawa couplings arise only through the coupling of the Higgs to the composite fermions, and their
mixings to the elementary fermions. 

In order to diagonalize the two-site lagrangian (\ref{eq:Ltotal}) one has to make a field rotation
from the elementary/composite basis to the mass eigenstate basis~\cite{Contino:2006nn}. 

The rotation in the fermionic sector can be conveniently parametrized in terms of the following mixing parameters
\begin{equation}
\tan\vphi_{tR} = \frac{\Delta_{R1}}{M_{\tilde{Q}*}} \equiv \frac{s_{R}}{c_{R}} , \qquad
\tan\vphi_{bR} =\frac{\Delta_{R2}}{M_{\tilde{Q}*}} \equiv \frac{s_{bR}}{c_{bR}} , \qquad
\tan\vphi_{L} =\frac{\Delta_{L}}{M_{Q*}}\equiv \frac{s_{1}}{c_{1}} , \qquad
\end{equation}

Here $\sin\vphi_{tR}$ (shortly indicated as $s_R$), $\sin\vphi_{bR}$ ($s_{bR}$), $\sin\vphi_{L}$ ($s_1$) respectively denote the degree of compositeness of  
$t_R$, $b_R$ and $(t_L, b_L)$. \\

The physical masses are given by

\begin{align}\label{eq:masses1}
\left\{\begin{array}{l}
m_{\tilde{T}}\\
m_{\tilde{B}}=m_{\tilde{T}} c_R / c_{bR}\simeq m_{\tilde{T}} c_R\\
m_{\tilde{T}5/3}=m_{\tilde{T}'5/3}=m_{\tilde{T}'}=m_{\tilde{B}'}=m_{\tilde{T}} c_R\\
m_{T}=m_B \\
m_{T2/3}=m_{T5/3}= m_T c_1 \end{array} \right. \ .
\end{align}
\noindent
More details can be found in App. \ref{TS10A}.\\

The final Yukawa lagrangian is shown in eq.(\ref{eq.Lagrange2_ts10}) of Appendix~\ref{TS10A}. 
We use an economical notation 
where symbols denoting elementary (composite) fields before the rotation now indicate the SM (heavy) fields.\\

After the EWSB, the SM top and bottom quarks acquire a mass, and the heavy masses in eqs.(\ref{eq:masses1}) get corrections 
of order $(vY_*/M_{Q*})^2$. In the following, we assume $r \equiv (vY_*/{ M_{Q*}}) \ll 1$ and compute all quantities at leading order in 
$r$.
We have the following expressions for the top and bottom masses:
\begin{equation}
 m_{t}=\frac{v}{2}Y_{*}s_{1}s_{R} \ , \ \  m_{b}=\frac{v}{\sqrt{2}}Y_{*}s_{1}s_{bR} \ .
\end{equation} 
$s_{bR}$ is forced to be very small, $s_{bR}\simeq 0$ ($c_{bR}\simeq 1$), in order for the model to reproduce the small ratio $m_b/m_t$ ($m_b/m_t\sim s_{bR}/s_R$).   \\

In the case of a right-handed top with a large degree of compositeness, $s_R\simeq 1$ ($c_R\ll 1$), we see from 
eq. (\ref{eq:masses1}) that heavy fermions in the $\mathcal{\tilde{Q}'}_{2/3}$ triplet as well the $\tilde{T}_{5/3}$ and the $\tilde{B}$ become much lighter than other fermionic resonances, $m_{\tilde{T}} c_R \ll m_{\tilde{T}} $. The same happens for $T_{5/3}$ and $T_{2/3}$ in the case of a composite left-handed top, $s_1\simeq 1$ ($c_1\ll 1$) . The appearance of relatively light resonances is tightly connected to the presence of the custodial symmetry in the composite sector; indeed, composite fermions which exhibit this behavior are generally known as custodians \cite{Contino:2006qr, Carena:2006bn, Agashe:2006at}.
In this analysis we will focus on the custodian $\tilde{B}$.  Indirect bounds on the $\tilde{B}$ mass from flavor observables and electroweak precision tests are model dependent. A specific study of $\Delta F=1$ constraints on the TS10 model considered in this analysis is presented in \cite{vignaroli_bsGamma}. In general, mass values roughly below $500$ GeV are allowed for this custodian heavy bottom (see also Ref. \cite{anastasiouRCM}). Moreover, also lighter $\tilde{B}$ could be easily accomodated if one considers particular assumptions on the flavor structure of the new strong sector, like minimal flavor violation or CP invariance \cite{Redi2011}. 
Mass values $m_{\tilde{B}}\lesssim 500$ GeV are thus realistic in composite models of the type considered here and still allowed by experimental constraints. Present searches for heavy vector-like quarks at the LHC \cite{Atlas_mBs_limit} put a still mild constraint, $m_{\tilde{B}}\gtrsim 420$ GeV, on the $\tilde{B}$ mass. \footnote{Ref. \cite{Atlas_mBs_limit} presents a search for pair production of heavy bottoms which decay predominantly as $\tilde{B}\to Wt$. We derive the limit $m_{\tilde{B}}\gtrsim 420$ GeV from \cite{Atlas_mBs_limit}, considering a value $BR(\tilde{B}\to Wt)=0.5$ instead of $BR(\tilde{B}\to Wt)=1$.}


\subsection{$\tilde{B}$ single-production and decays } 

%

The interactions beyond the SM which are relevant for our analysis of the $\tilde{B}$ single production and subsequent decay into a composite Higgs can be restricted to the following terms: 

 \begin{align}  \label{eq:LBs}
\begin{split} 
 {\cal L}_{\tilde{B},h}  =
  & \bar{\tilde{B}}\left(i\dslash -m_{\tilde{B}}\right)\tilde{B} + \frac{1}{2} \,\text{Tr}\left\{  \partial_\mu {\cal H}^\dagger \partial^\mu {\cal H} \right\} - V( {\cal H}^\dagger {\cal H}) \\[0.2cm]
  & -Y_{*}s_{1}c_{bR}\left(\bar{b}_{L}\phi_{0}\tilde{B}_{R}+\bar{t}_{L}\phi^{+}\tilde{B}_{R}\right) + h.c.
\end{split} 
\end{align}

\begin{figure}[t]
\subfigure[t][$\tilde{B}$ single production.]{\includegraphics[width=0.25\textwidth, trim=0cm 0cm 0cm 4cm, clip=false]{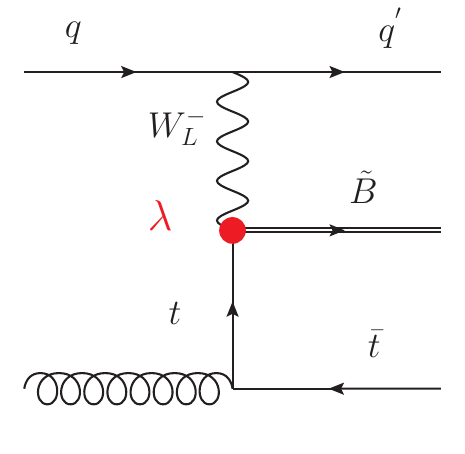}\label{Bsprod_fig}}\hspace{1cm}
\subfigure[][cross section values for the $\tilde{B}t+X + c.c.$ single production (we consider both the $\tilde{B}\bar{t}+X$ and the $\bar{\tilde{B}}t+X$ channels) at the LHC with $\sqrt{s}=14$ TeV and $\sqrt{s}=8$ TeV. We set the $\lambda$ coupling to the reference value $\lambda=3$; cross sections scale with $\lambda$ as $\lambda^2$.]{\includegraphics[width=0.65\textwidth]{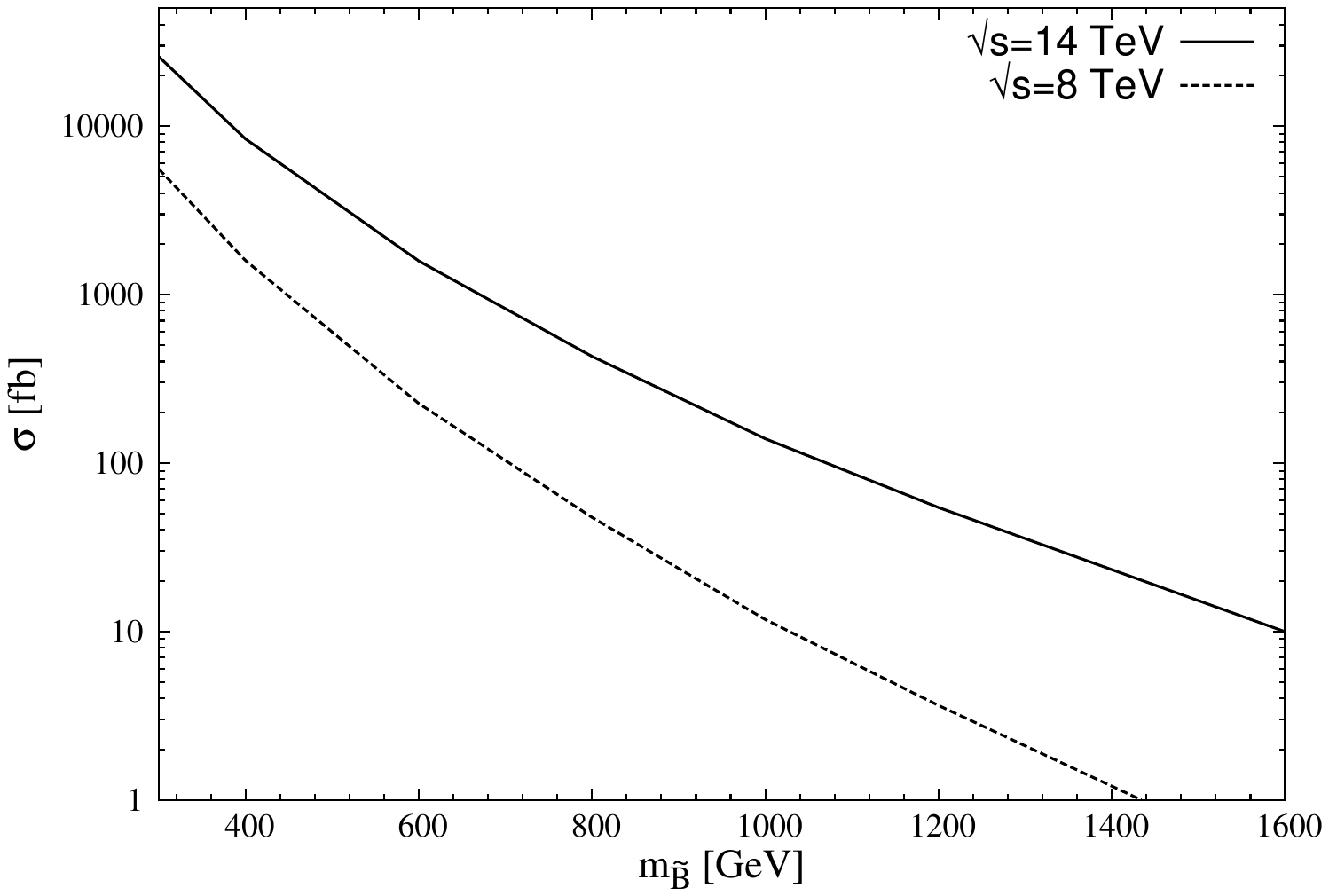}\label{Bsprod_xsec}}
\caption{\textit{$\tilde{B}$ single production.}}
\end{figure}

The $\tilde{B}$ branching ratios (BR) are essentially fixed by the equivalence theorem to be $BR(\tilde{B}\to W_L t)\simeq 0.50$,  
$BR(\tilde{B}\to Z_L b)\simeq BR(\tilde{B}\to hb)\simeq 0.25$. The rate for the $\tilde{B}$ decays can be found in Appendix \ref{BRsHf}. They depend quadratically on the vertex
\begin{equation}\label{eq:lambda}
\lambda= Y_{*} s_1 c_{bR}\simeq Y_{*} s_1\ ,
\end{equation}
which can be directly read from the lagrangian (\ref{eq:LBs}). We remind that $Y_{*}$ represents the Yukawa coupling among composite states (see eq. (\ref{eq:Lcomp})). The last equality in eq. (\ref{eq:lambda}) comes from the fact that 
$s_{bR}\simeq 0$ ($c_{bR}\simeq 1$), to reproduce the small ratio $m_b/m_t$.
The $\lambda$ coupling of the heavy bottom $\tilde{B}$ with the electro-weak bosons and the SM quarks affects both the decay and the single-production of the $\tilde{B}$. The single production is described by the diagram in Figure \ref{Bsprod_fig}, where the intermediate exchange of a $W_L$ ($Z_L/h$) mediates the production of a $\tilde{B}$ in association with a top (bottom) and a light quark. 
In particular, we will focus on the final state $\tilde{B}t+jets$, where the $\tilde{B}$ subsequently decays into a Higgs plus a bottom and the top decays leptonically. We also consider the decay of the Higgs into a $b\bar{b}$ pair, with the same BR of that of a SM Higgs of $125$ GeV. 

Despite we are referring to a specific composite Higgs model, our analysis could be easily extended to other scenarios, considering that it depends on just two parameters (once we fix the Higgs mass and the $h\to b\bar{b}$ branching ratio): 
\begin{equation}
\lambda \ , \qquad m_{\tilde{B}} \ .
\end{equation}  
 
 At the end we will analyze the LHC sensitivity on the $(\tilde{B}\to hb)t+jets$ channel in the $(m_{\tilde{B}}, \lambda)$ parameter space. We now fix $\lambda=3$ and we consider several $\tilde{B}$ mass values. In the specific case of TS10, $\lambda=3$ could be realized, for example, if we have a Yukawa coupling $Y_{*}\sim 3$ (which is a reference value generally considered in composite Higgs models) and a left-handed top with a large degree of compositeness, $s_1\sim 1$. \\
The precise value for the $\tilde{B}$ BRs and width can be derived from Fig. \ref{fig:Bs-decay} in Appendix \ref{TS10A}; in the range of mass we will consider and for the reference value $\lambda=3$ we have 
\begin{equation} 
BR(\tilde{B}\to hb)\simeq 0.26  \qquad \Gamma(\tilde{B})/m_{\tilde{B}}\simeq 0.16
\end{equation}


Fig. \ref{Bsprod_xsec} shows the cross section for the $\tilde{B}t+X + c.c.$ single production at the LHC with $\sqrt{s}=14$ TeV and $\sqrt{s}=8$ TeV, as a function of the $\tilde{B}$ mass.  
In Fig. \ref{Bsprod_xsec} we consider both the $\tilde{B}$ charges. Notice that, due to the different content of the up and down partons in the proton, the cross section for the single-production of $\bar{\tilde{B}}$ is roughly 2 times that for $\tilde{B}$. We find $\sigma(\bar{\tilde{B}})/\sigma(\tilde{B})\simeq 2$ at 14 TeV LHC and $\simeq2.4$ at 8 TeV LHC (these numbers slightly grow up with the increasing of the $\tilde{B}$ mass). In our analysis we will exploit mainly kinematic cuts and we will not make use of this charge asymmetry. Anyway it could represent a promising variable to discriminate between signal and background, especially at 8 TeV.


\section{Analysis}
\label{sec:analysis}

In this section we discuss the prospects of observing the composite Higgs production in the channel $pp\to (\tilde{B}\to (h\to bb)b)t+X$ at the LHC. We consider the leptonic decay of the top. The physical final state is thus:
\begin{equation} \label{eq:finalstate}
pp \to l^\pm \! + n\, jets \, +  \not\!\! E_T\ .
\end{equation}
We will present a simple parton-level analysis aimed at assessing the LHC discovery reach. 
We consider two center-of-mass energies: $\sqrt{s}=8\,$TeV,
the energy of the current phase of data taking, and $\sqrt{s}=14\,$TeV, the design energy that will
be reached in the second phase of operation of the LHC. Our selection strategy and the set of kinematic cuts that we will design do not depend, however, on the value of the collider energy. This is because they will be optimized to exploit the peculiar kinematics of the signal, and a change in the collider energy mainly implies a rescaling of the production cross sections of signal and background via the parton luminosities, without affecting the kinematic distributions.

\subsection{Montecarlo simulation of signal and background}

We simulate the signal by using MadGraph v4~\cite{MG-ME}, after implementing the TS10 model with Feynrules \cite{Feynrules}, 
while for the background we make use of both MadGraph and ALPGEN~\cite{Mangano:2002ea}.~\footnote{
The factorization and renormalization scales have been set to be equal and chosen as follows:
$Q= m_{\tilde{B}}/4$ for the signal; $Q = \sqrt{m_W^2 + \sum_j p_{T j}^2}$ for $WWbb+jets$; 
$Q = \sqrt{m_W^2 +  p_{T W}^2}$ for $Wbb+jets$ and $W+jets$.
}
In our parton-level analysis jets are identified 
with the quarks and gluons from the hard scattering.
If two quarks or gluons are closer than the separation $\Delta R =0.4$, they are merged into a single jet whose four-momentum is the
vectorial sum of the original momenta.
We require that the jets and the leptons satisfy 
the following set of acceptance and isolation cuts:
\begin{equation}
\begin{aligned}
\ptj &\geq 30\, \gev  & \quad  | \etaj | &\leq 5    & \quad 
 \Delta R_{jj} &\geq 0.4 \\[0.2cm]
\ptl &\geq 20\, \gev & \quad  | \etal | &\leq 2.5  & \quad 
 \Delta R_{jl} &\geq 0.4 \, .
\end{aligned}
\label{eq:acceptance}
\end{equation}
Here $\ptj$ ($\ptl$) and $\etaj$ ($\etal$) are respectively 
the jet (lepton) transverse momentum and pseudorapidity, and 
$\Delta R_{jj}$, $\Delta R_{jl}$ denote the jet-jet and jet-lepton separations.

Detector effects are roughly accounted for by performing a simple Gaussian smearing on the 
jet energy and momentum absolute value with $\Delta E/E = 100\%/\sqrt{E/\text{GeV}}$, and on the jet momentum direction using
an angle resolution $\Delta\phi =0.05$ radians and $\Delta\eta = 0.04$. Moreover, the missing energy $\etmiss$ of each event has been computed
by including a Gaussian resolution $\sigma(\etmiss) = a \cdot \sqrt{\sum_i E^i_T/\gev}$, where $\sum_i E^i_T$ is the scalar sum of the transverse
energies of all the reconstructed objects (electrons, muons and jets). We choose $a=0.49$.~\footnote{This numerical value, as well as the $b$-tagging efficiency and rejection rate and the resolution parameters considered in the jet smearing, have been chosen according to the performance of the ATLAS detector~\cite{Aad:2008zzm}.} 

Table~\ref{tab:fractions} shows the fraction of signal events
where the jet content is reconstructed to be respectively $5j$ (where $j$ denotes either a light jet or a $b$-jet), $4j$ 
and $3j$, for different values of the $\tilde{B}$ mass and for $\sqrt{s}=14\,$TeV (very similar numbers hold for $\sqrt{s}=8\,$TeV). The numbers in parenthesis in the $4j$ column represent the percentage of the events, among those with four jets, where the $b\bar b$ pair from the Higgs merges into a single fat jet (which we denote as $Mj$).
%
\begin{table}
\begin{center}
\begin{tabular}{c|ccc}
  $m_{\tilde{B}}$ & $5j$ & $4j$  $(3j+1Mj)$ & $3j$ \\[0.05cm]
\hline
  & & & \\[-0.3cm]
   $0.4\,$TeV &  0.53 & 0.38 (13\%) & 0.08  \\[0.1cm]
   $0.6\,$TeV &  0.56 & 0.37 (13\%)  & 0.07  \\[0.1cm]
   $0.8\,$TeV & 0.57 & 0.36  (16\%) & 0.07  \\[0.1cm]
   $1.0\,$TeV & 0.57 &  0.36 (19\%) &  0.07\\[0.1cm]
   $1.2\,$TeV & 0.55 & 0.37 (26\%)  &  0.08
\end{tabular}
\caption{\small 
Fraction of  signal events
where the jet content is reconstructed to be respectively $5j$ (where $j$ denotes either a light jet or a $b$-jet), $4j$ and $3j$, for different values of the $\tilde{B}$ mass and $\sqrt{s}=14\,$TeV. The numbers in parenthesis in the third column represent the percentage of the events with four jets where one of the four jets is a fat jet ($Mj$) resulting from the merging of the $b\bar{b}$ pair from the Higgs.
\label{tab:fractions}
}
\end{center}
\end{table}
%
%

The events with $4j$ are mainly constituted by events where one jet is soft, with $p_T<30$ GeV. Events where one jet is a merged jet increase with larger masses, because the Higgs is more boosted in this case, but the percentage is not so relevant; sophisticated technique of tagging boosted object could not be useful in this case. 
Tab. \ref{tab:fractions} shows that the fraction of events with four jets is not negligible.
For this reason in our analysis we will select events with at least four jets and exactly one lepton. We further require the $b$-tagging of at least two $b$-jets. %

\begin{equation} \label{eq:evsel}
pp \to l^\pm \! + n\, jets \, +  \not\!\! E_T\, , \qquad n\geq 4 \, , \ \ \ \text{At least 2 $b$-tag}
\end{equation}
where all objects must satisfy the acceptance and isolation cuts of eq.(\ref{eq:acceptance}). 

The largest SM background after the event selection of eq.(\ref{eq:evsel}) is the irreducible background $WWbb+jets$, 
which includes the resonant sub-processes $Wtb+jets$ (single top) and $t\bar t +jets$.
The latter, in particular, gives the largest contribution.
Other backgrounds which will turn out to be relevant after imposing our full set of kinematic cuts are $Wbb+jets$ and $W+jets$, where at least two of the light jets are mistagged as $b$-jets . \\
For each event of the samples with $b$ quarks (with only light jets), we estimate, taking into account a binomial distribution, the efficiency $\eps_{AL2b}$ ($\zeta_{AL2b}$) of tagging at least two $b$-jets as: 
\begin{equation} \label{eq:btag}
\eps_{AL2b} =\sum_{2\leq k \leq n_b} \left( \begin{array}{cc} n_b\\ k \end{array}\right) \left( \epsilon_b \right)^{k}\left(1-\epsilon_b\right)^{n_b-k}, \qquad \zeta_{AL2b}=\left( \begin{array}{cc} n_j\\ 2 \end{array}\right)\zeta^2_b \ ,
\end{equation}
where $\eps_b = 0.6$ is the efficiency of the $b$-tag and $\zeta_b = 0.01$ is the probability of mistagging a light jet as a $b$-jet. $n_b$ is the number of $b$ quarks with $|\eta_b|<2.5$, $n_j$ is the number of light jet with $|\eta_j|<2.5$. We found efficiencies of about 0.81 for the signal events (which have typically four central $b$ quarks), of about 0.35 for the $WWbb+jets$ and the $Wbb+jets$ background (with typically two central $b$ quarks) and of about $4\cdot 10^{-4}$ for the $W+jets$ background.

We have simulated the $WWbb$ events by using MadGraph, while the other backgrounds are generated with ALPGEN.
For simplicity, in our analysis we include all the samples with increasing multiplicity of light jets in the final state. 
This is a redundant procedure which could lead to a double counting of kinematic configurations.
A correct procedure would be resumming soft and collinear emissions by means of a parton shower,
and adopting some matching technique to avoid double counting. However, retaining all the $W+ n\, jets$ samples, we expect to obtain a conservative estimate
of the background. Moreover some of the cuts we will impose tend to suppress
the events with larger number of jets and thus to reduce the amount of double counting.

%
%

\subsection{Reconstruction of the heavy bottom and of the Higgs resonances}\label{sec:reco}

We are focusing our analysis on the channel $pp\to (\tilde{B}\to (h\to bb)b)t+X$. 
The physical final state is that in eq. (\ref{eq:evsel}).

We have $n\geq 4$ jets in the final state; two of these jets should not come from $\tilde{B}$, one is the $b$-jet from the top decay, the other is the light jet from the initial parton that emits the intermediate $W$. In order to reconstruct the $\tilde{B}$ resonance, we need to tag these two jets and discard them from the possible constituents of the heavy-bottom resonance.\\

%
At a first step we tag the top and its decay products. 
The procedure we adopt requires the reconstruction of the momentum of the neutrino.
The transverse momentum of the neutrino can be reconstructed from the transverse missing momentum;
this latter can be estimated, considering a $p^{TOT}_T=0$ hypothesis, as $p^{miss}_T=-\sum p_T$, where $\sum p_T$ is the sum over the $p_T$ of all the detected final states.
Once we have estimated the neutrino transverse momentum, we can derive the neutrino longitudinal momentum, $p_z$, by requiring that the
neutrino and the lepton reconstruct an on-mass-shell $W$, $M_{l\nu} = 80.4$ GeV. The condition
\begin{equation}
 (E^{l}+E^{\nu})^{2}-(p^{l}_x+p^{\nu}_x)^{2}-(p^{l}_y+p^{\nu}_y)^{2}-(p^{l}_z+p^{\nu}_z)^{2}=M^2_W
\label{neutrino}
\end{equation}
 gives two solutions for $p^{\nu}_z$. 
 We find that in the $\simeq 20 \%$ of the events, both for the signal and the background, the eq. (\ref{neutrino}) has imaginary solutions 
(this corresponds to the case of a quite off-shell leptonically decayed $W$). In this case we decide to throw out the event. 
Our neutrino reconstruction procedure has, therefore, an efficiency of about the $80\%$. 
Once we have reconstructed the momentum of the neutrino, we want to reconstruct the top which is in our signal 
and, in particular, to tag the jet associated to its decay. 
To do this, we first reconstruct the leptonically decayed $W$ and then we consider all the possible $Wj$ combinations. 
In each event we have two $W\to l\nu$ candidates, one for each of the two solutions of the neutrino longitudinal momentum, eq.(\ref{neutrino}).
The $Wj$ pair that gives the $M_{Wj}$ invariant mass closest to the top mass, $m_t=174$ GeV, is selected as the pair coming from the decay of the top. Notice that this procedure allows, as a bonus, to fully reconstruct the neutrino.
The top 4-momentum is then reconstructed by summing on the 4-momentum of the $W$ and of the $j$ that form the selected pair. At this stage of the analysis, we do not want yet to distinguish between the signal and the background. Our aim is to recognize the top we know to be in the signal.
Therefore, we choose to keep all the events in a $M_{top}$ wide range, $[150$ GeV, $230$ GeV$]$, in order to catch the most part of the signal events. \\
We thus proceed to tag the light jet from the parton that emits the intermediate $W$. 
We can do this by considering that this jet tends to be emitted at very high rapidity. 
As also discussed in \cite{Mrazek:2009yu} and first found in \cite{Willenbrock}, the intermediate $W$ tends to carry only a small fraction of the initial parton energy, in order to maximize its propagator. At the same time, it must have enough energy to produce the heavy bottom. Thus, the quark in the final state that originates from the parton emitting the $W$ has  a high energy and a small transverse momentum (we find a ratio of about ten between the quark energy and the quark transverse momentum). This results in a final light jet with high rapidity, $|\eta|\gtrsim$ 2.5. This is a peculiarity of the topology of the signal that we exploit to reconstruct the $\tilde{B}$ resonance and that we will also further exploit to discriminate between the signal and the background.
We check from the Montecarlo simulation that the light jet in the signal represents the jet (not coming from the top) with the highest rapidity in about the $80-90 \%$ of the cases, the fraction is 0.80 for $m_{\tilde{B}}=0.4$ TeV and grows up to 0.89 for a $\tilde{B}$ of $1.2$ TeV. We thus tag the light-jet, by assuming that it coincides with the jet (not coming from the top) with the highest rapidity.\\
The heavy-bottom is finally reconstructed by summing on the 4-momentum of all the jets in the final state except the jet coming from the reconstructed top and the tagged light-jet.\\
The Higgs can be reconstructed by considering that it is produced from the decay of the $\tilde{B}$ 
in association with a bottom. This latter can be distinguished from the jets produced from the Higgs decay, by the fact that it tends to be more energetic than the other jets, being it directly produced from the decay of the heavy fermion. 
Therefore, we reconstruct the Higgs by summing on the 4-momentum of all the jets from the heavy-bottom, with the exception of the most energetic jet among them.\\

The reconstruction of the $\tilde{B}$ and Higgs resonance is crucial for the discovery of such particles and to obtain an estimate value of their masses. The reconstruction of the intermediate final state $\tilde{B}tj$ (see Figure \ref{Bsprod_fig}) is also very useful to design a strategy for the reduction of the background.

\subsection{Event selection}

In Table \ref{tab:cutflow1} we report the value of the cross section for the signal and the 
main SM backgrounds after the selection (\ref{eq:evsel}) based on the acceptance and isolation cuts of eq.(\ref{eq:acceptance}) and the $b$-tag efficiencies of (\ref{eq:btag}) and after the neutrino and top reconstruction, for $\sqrt{s} = 8\,$TeV and $\sqrt{s} = 14\,$TeV.

%
\begin{table}
\begin{center}
{\small
\begin{tabular}{|c|cc|cc|}
\multicolumn{1}{c}{} & \multicolumn{2}{c}{\textsf{LHC 8 TeV}} &   \multicolumn{2}{c}{\textsf{LHC 14 TeV}}\\[0.1cm]
\hline
 & \multicolumn{1}{c}{\textsf{acceptance}} & \multicolumn{1}{c|}{$\nu$+top reco} & \multicolumn{1}{c}{\textsf{acceptance}} & \multicolumn{1}{c|}{$\nu$+top reco} \\[0.1cm]
 & & & & \\[-0.3cm]
Signal $m_{\tilde{B}}=0.4\,$TeV & 30.8 & 24.3 & 235 & 136 \\[0.25cm]
 Signal $m_{\tilde{B}}=0.6\,$TeV & 4.42 & 3.43 & 37.2 & 23.7 \\[0.25cm]
 Signal $m_{\tilde{B}}=0.8\,$TeV & 1.01 & 0.773 & 9.63 & 6.46 \\[0.25cm]
 Signal $m_{\tilde{B}}=1.0\,$TeV & & & 3.11 & 2.18 \\[0.25cm]
 Signal $m_{\tilde{B}}=1.2\,$TeV & & & 1.14 & 0.850 \\[0.35cm]
  $WWbb$            & 3510 & 2840  & 16700 & 13100\\[0.25cm]
  $WWbbj$            & 2160 & 1730 &  10600 & 8510\\[0.25cm]
  $WWbbjj$            & 800 & 633 & 4640 & 3560\\[0.25cm]
  $Wbbjj$                & 137  & 93.0 & 573 & 375 \\[0.25cm]
  $Wbb3j$                &  52.9  & 36.2 & 324 & 213 \\[0.25cm]
  $W4j$              &  11.2  &  7.61 & 38.4 & 25.0\\[0.25cm]
  $W5j$             &  4.42  &  3.05 & 18.7 & 12.3 \\[0.35cm]
  Total                  & & & &\\
  background       & 6680 &  5350 & 33000 & 25800  \\[0.15cm]
\hline
\end{tabular}
}
\caption{
\label{tab:cutflow1}
\small 
Cross sections, in fb, for the signal (with $\lambda=3$) and the main backgrounds after the selection (\ref{eq:evsel}) based on the acceptance cuts of eq.(\ref{eq:acceptance}) and the $b$-tag efficiencies of (\ref{eq:btag}) and
after the reconstruction of the neutrino and of the top quark, for $\sqrt{s} = 8\,$TeV and $\sqrt{s} = 14\,$TeV.
}
\end{center}
\end{table}

\begin{figure}[tbp]
\begin{center}
\includegraphics[width=0.485\textwidth,clip,angle=0]{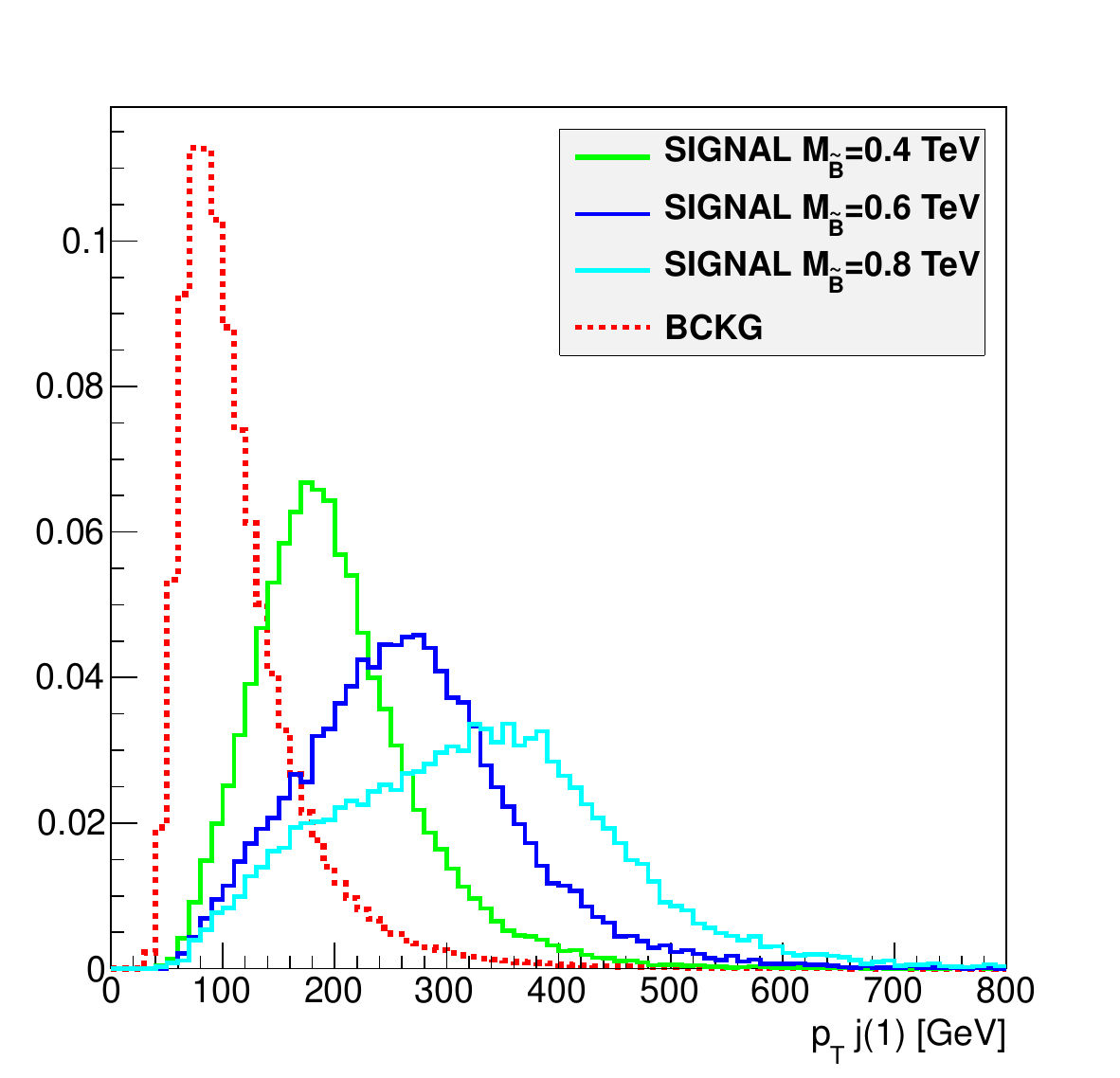}
\includegraphics[width=0.485\textwidth,clip,angle=0]{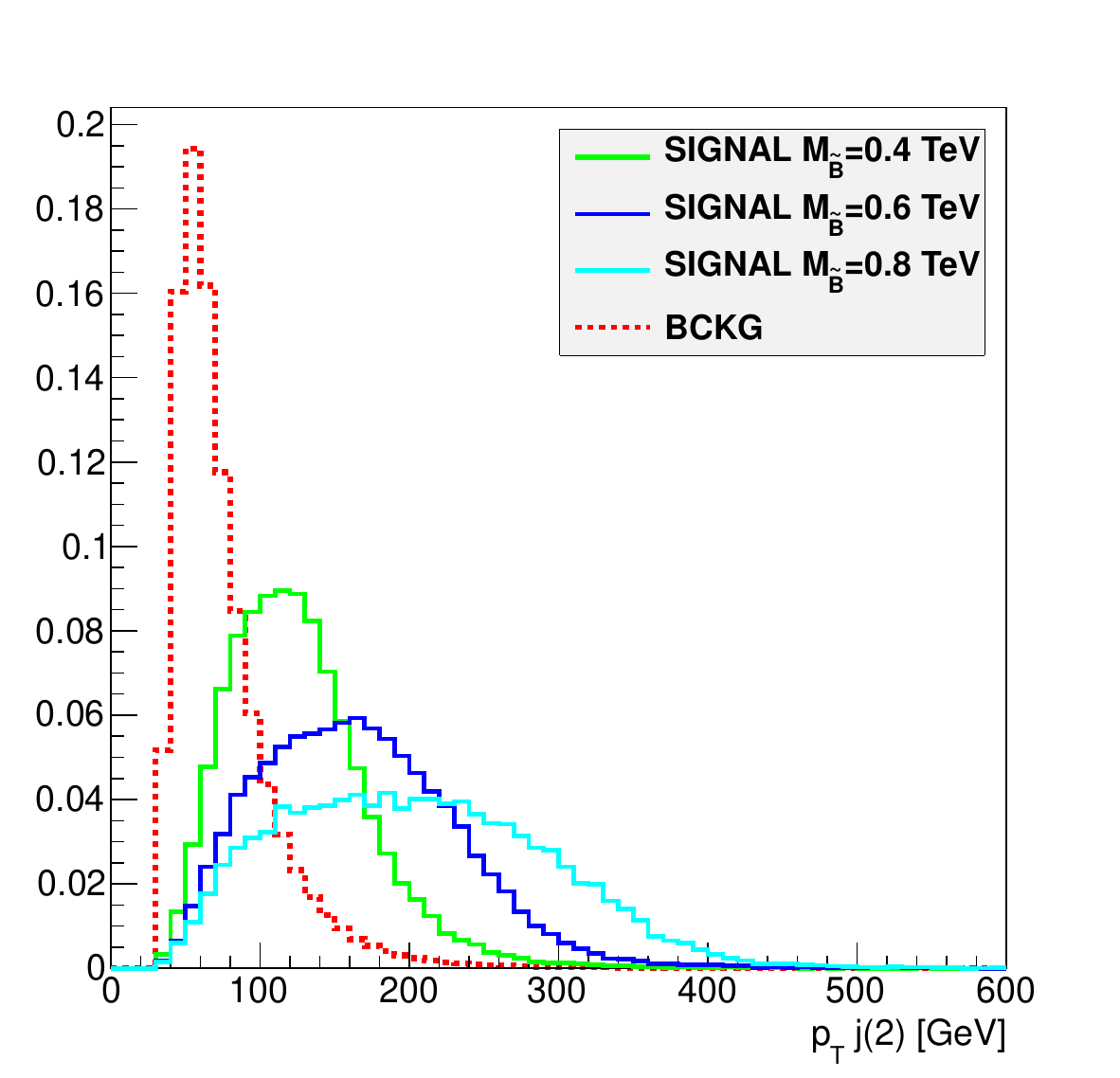}
\\[0.3cm]
\includegraphics[width=0.485\textwidth,clip,angle=0]{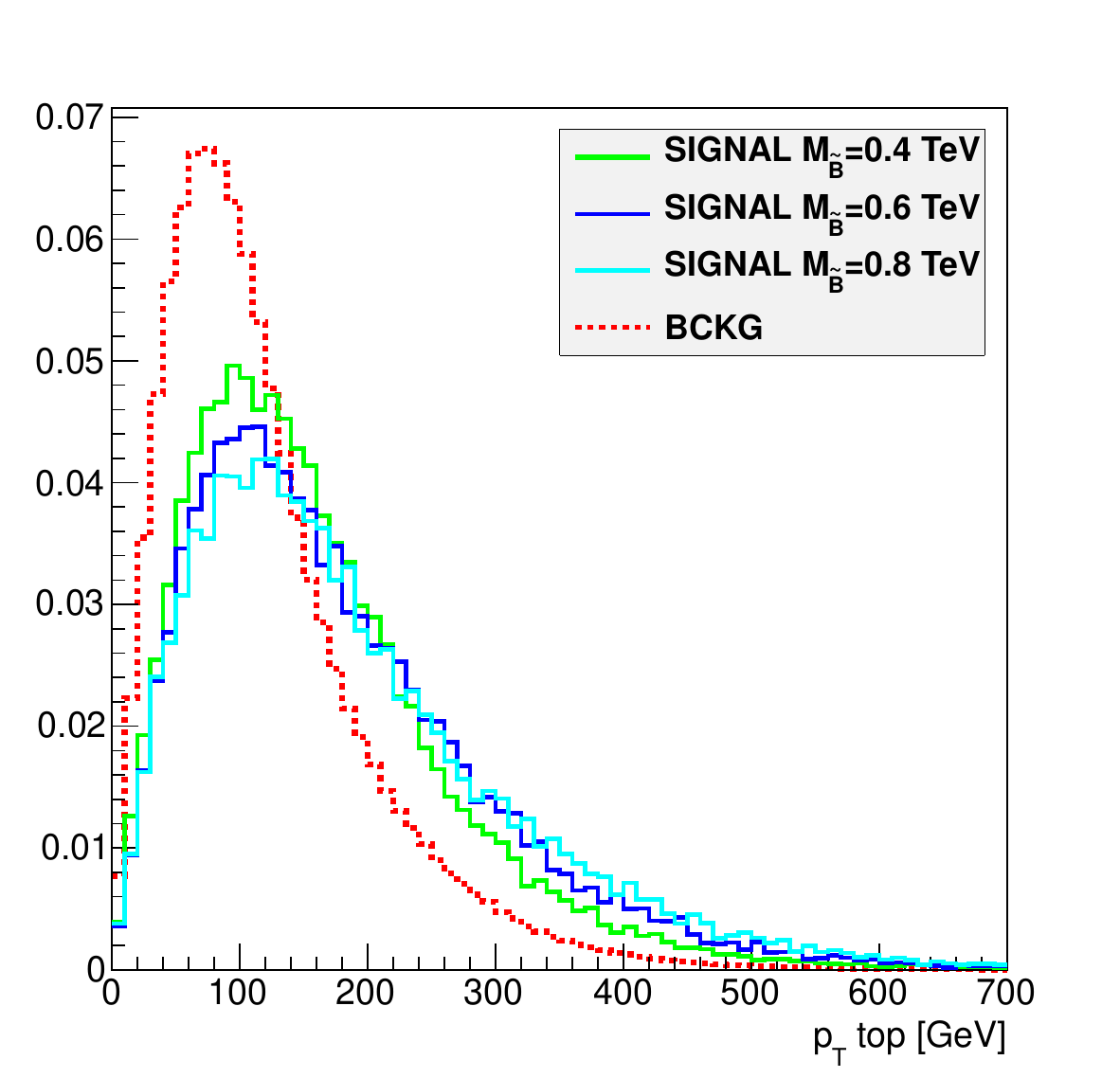}
\includegraphics[width=0.485\textwidth,clip,angle=0]{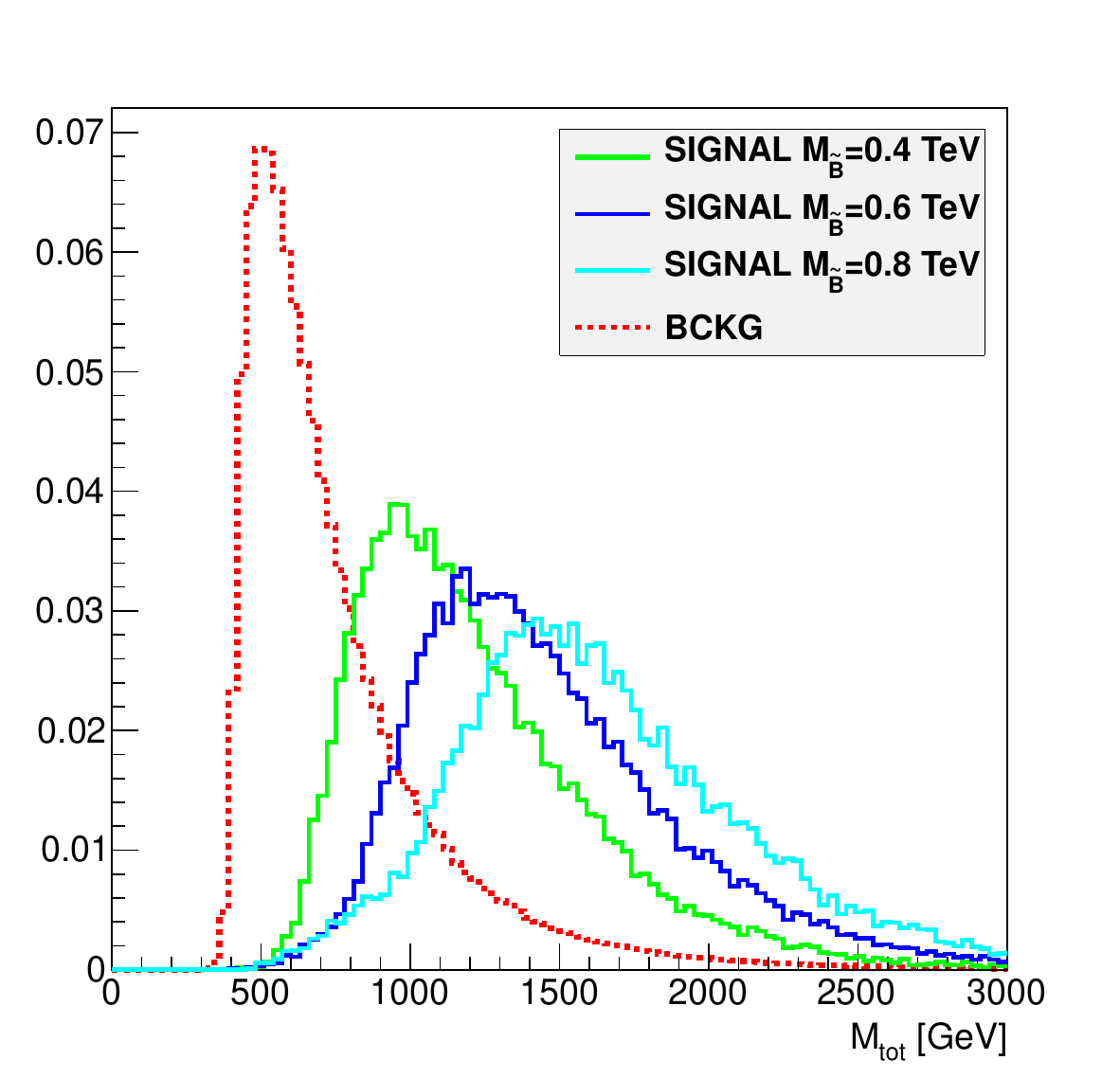}
\caption[]{
\label{pT_distribution}
\small
Differential distributions after the neutrino and  top
reconstruction 
for $\sqrt{s} =8\,$TeV.
Upper left plot: $p_T$ of the hardest jet, not coming from the top, $p_{T} j(1)$; Upper right plot: $p_T$ of the second hardest jet, not coming from the top, $p_{T}j(2)$; Lower left plot: $p_T$ of the reconstructed top;
Lower right plot: invariant mass of the system $\tilde{B}$ plus top plus jets, where the $\tilde{B}$ and the top are the objects we have reconstructed following the procedure explained in Sec. \ref{sec:reco}, $M_{tot}$.
The continuous lines show the signal at different $\tilde{B}$ mass values, the dashed (red) line
shows the total background.
All the curves have been normalized to unit~area. 
}
\end{center}
\end{figure}

\begin{figure}[tbp]
\begin{center}
\includegraphics[width=0.485\textwidth,clip,angle=0]{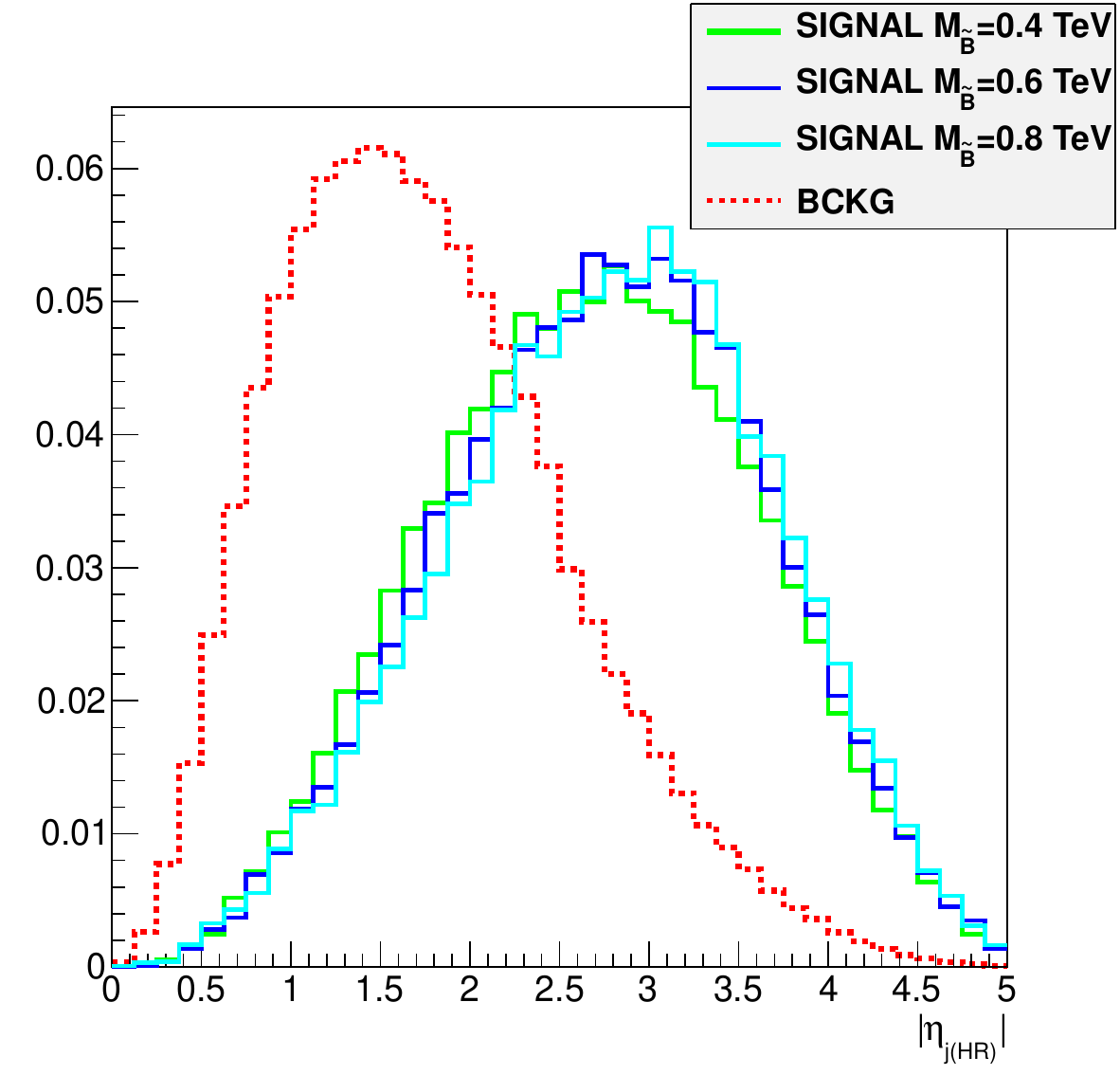}
\caption[]{
\label{MRj_distribution}
\small
Differential distribution of the rapidity of the jet with the highest rapidity ($j (HR)$), after the neutrino and  top
reconstruction 
for $\sqrt{s} =8\,$TeV.
The continuous lines show the signal at different $\tilde{B}$ mass values, the dashed (red) line
shows the total background.
All the curves have been normalized to unit~area. 
}
\end{center}
\end{figure}

\begin{figure}[tbp]
\begin{center}
\includegraphics[width=0.485\textwidth,clip,angle=0]{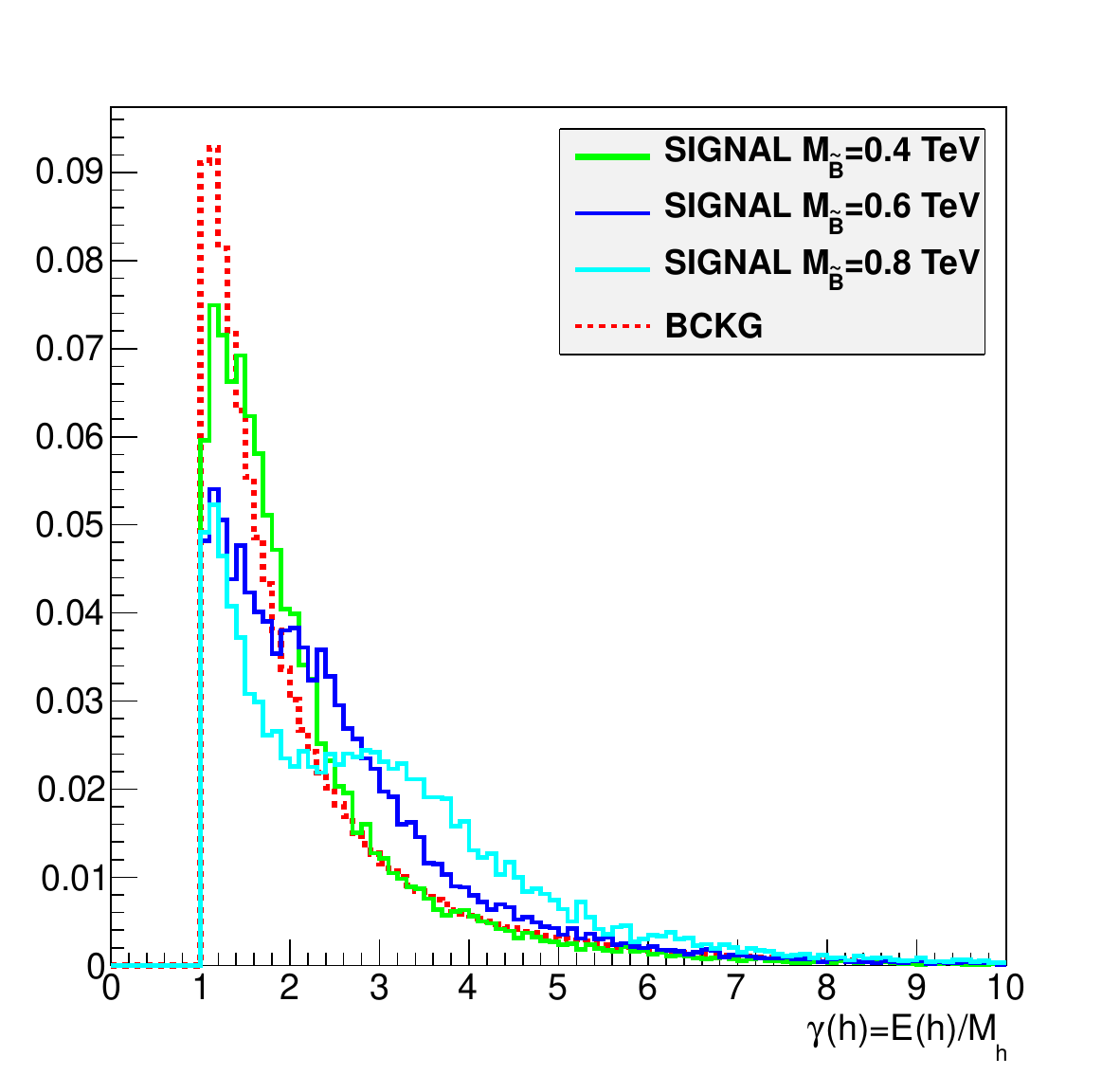}
\includegraphics[width=0.485\textwidth,clip,angle=0]{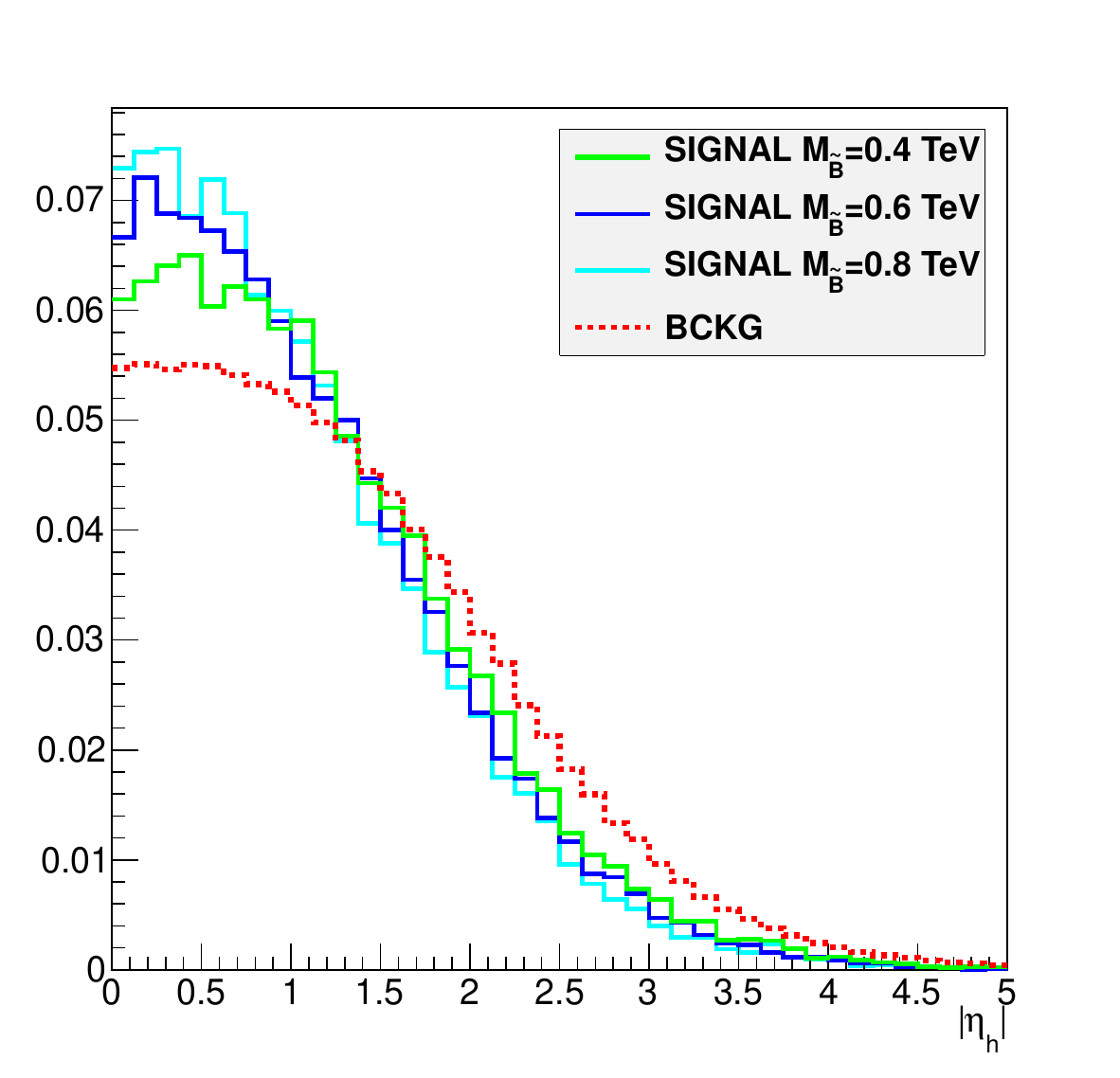}
\caption[]{
\label{higgs_distribution}
\small
Differential distributions after the neutrino and  top
reconstruction 
for $\sqrt{s} =8\,$TeV.
Left plot: distribution of the Lorentz factor of the reconstructed Higgs, $\gamma (h)$; 
Right plot: distribution of the Higgs rapidity, $|\eta_h|$.
The continuous lines show the signal at different $\tilde{B}$ mass values, the dashed (red) line
shows the total background.
All the curves have been normalized to unit~area. 
}
\end{center}
\end{figure}

One can see that at this stage the background dominates by far over the signal.
We can however exploit the peculiar kinematics of the signal to perform a set of cuts that reduce the background to 
a much smaller level. 
One of the peculiarities of the signal is the presence of the heavy fermionic resonance among the intermediate final states. Its production requires the exchange of a large amount of energy and leads to very energetic final states. We find very effective applying a cut on the invariant mass of all the $\tilde{B}tj$ particles in the intermediate final state as well a cut on the transverse momentum of the jet with the highest transverse momentum, among the final jets that do not come from the top. We will also apply a cut on the $p_T$ of the second hardest jet (not coming from the top) and on the $p_T$ of the top. The other important characteristic of the signal topology, as already discussed, is the presence of a final jet at very high rapidity. 
We exploit this feature imposing a cut on the rapidity of the most forward jet. Further conditions are imposed on the Higgs, which is slightly boosted, since it comes from a heavy particle, and is emitted at quite low rapidity.\\
We show in Fig. \ref{pT_distribution} the $p_T$ distributions of the hardest ($j(1)$) and of the second hardest ($j(2)$) jet, not coming from the top, the $p_T$ distribution of the top and the $M_{tot}$ invariant mass of the system $\tilde{B}tj$, where the $\tilde{B}$, the top and the light-jet are the objects in the intermediate final state, which we have reconstructed following the procedure explained in Sec. \ref{sec:reco} .  
In Fig. \ref{MRj_distribution} we show the $|\eta|$ distribution of the jet with the highest rapidity ($j  (HR)$), in Fig. \ref{higgs_distribution} we report the distribution of the Lorentz factor of the reconstructed Higgs and of the Higgs rapidity.
We show the distributions (normalized to unit area) for the total background and for the signal referred to different $\tilde{B}$ mass values.\\
As expected, the signal from highest $\tilde{B}$ mass values has ever more energetic final particles; especially the distributions on $p_{T}j(1)$ and on $M_{tot}$ shift on larger values with the increasing of $m_{\tilde{B}}$.
We find a set of optimized cuts that minimizes the integrated luminosity needed for a $5\sigma$ discovery of the signal with $m_{\tilde{B}}=400$ GeV. In the cases of higher $\tilde{B}$ masses we will refine in a second step the cuts on $M_{tot}$ and on $p_{T}j(1)$.
We define the discovery luminosity to be the integrated luminosity for which a goodness-of-fit test of the SM-only hypothesis with Poisson 
distribution gives a  $\text{p-value}=2.85\times 10^{-7}$, which corresponds to a $5\sigma$ significance in the limit of a gaussian distribution. \\
These optimized cuts are:

\begin{equation} \label{eq:optimized}
\begin{aligned}
& p_{T}\ j(1)>170\ \text{GeV} && p_{T}\ j(2)>100\ \text{GeV} && p_{T}\ top >100\ \text{GeV} && M_{tot}>1.2 \ \text{TeV}\\[0.4cm]
& |\eta_{j(HR)}|>2.5 && p_T\ j(HR)>30\ \text{GeV} \\[0.4cm]
& \gamma (h)>1.4 && |\eta_{h}|<1.8 && M_{h-j}< 70\ \text{GeV} .
\end{aligned}
\end{equation}

The cut $p_T j(HR)>30$ GeV is imposed in order to avoid possible conflicts with the jets from initial state radiation\footnote{We checked that this cut can be also enhanced to $40$ GeV without changing significantly the results.}; 
$M_{h-j}$ represents the invariant mass of the objects which form the reconstructed Higgs, from which the most energetic jet (among them) is subtracted. The cut on $M_{h-j}$ reduces particularly the backgrounds with more than $5$ jets in the final state.\\
  
\begin{figure}
\includegraphics[width=0.485\textwidth,clip,angle=0]{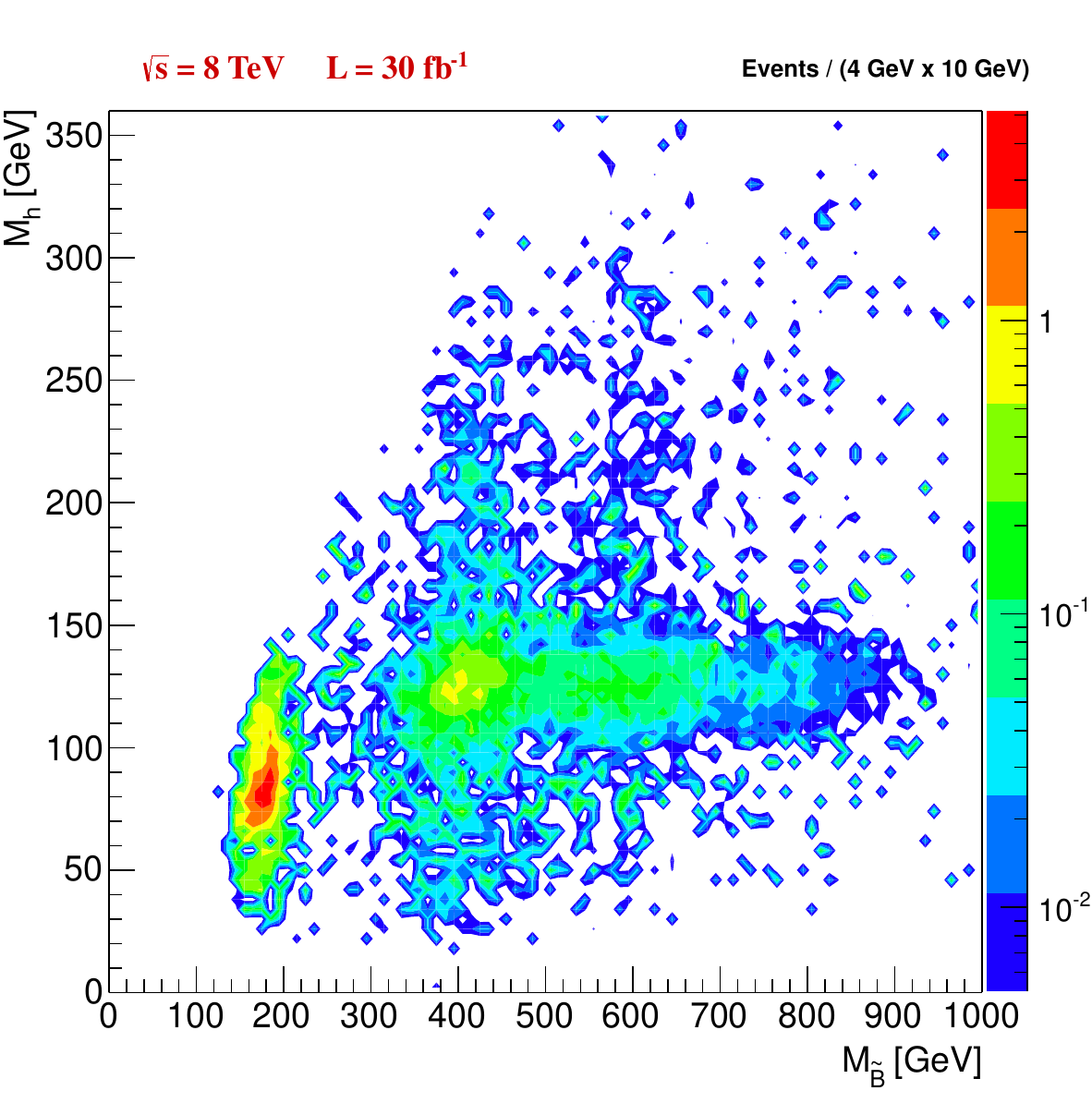}
\includegraphics[width=0.485\textwidth,clip,angle=0]{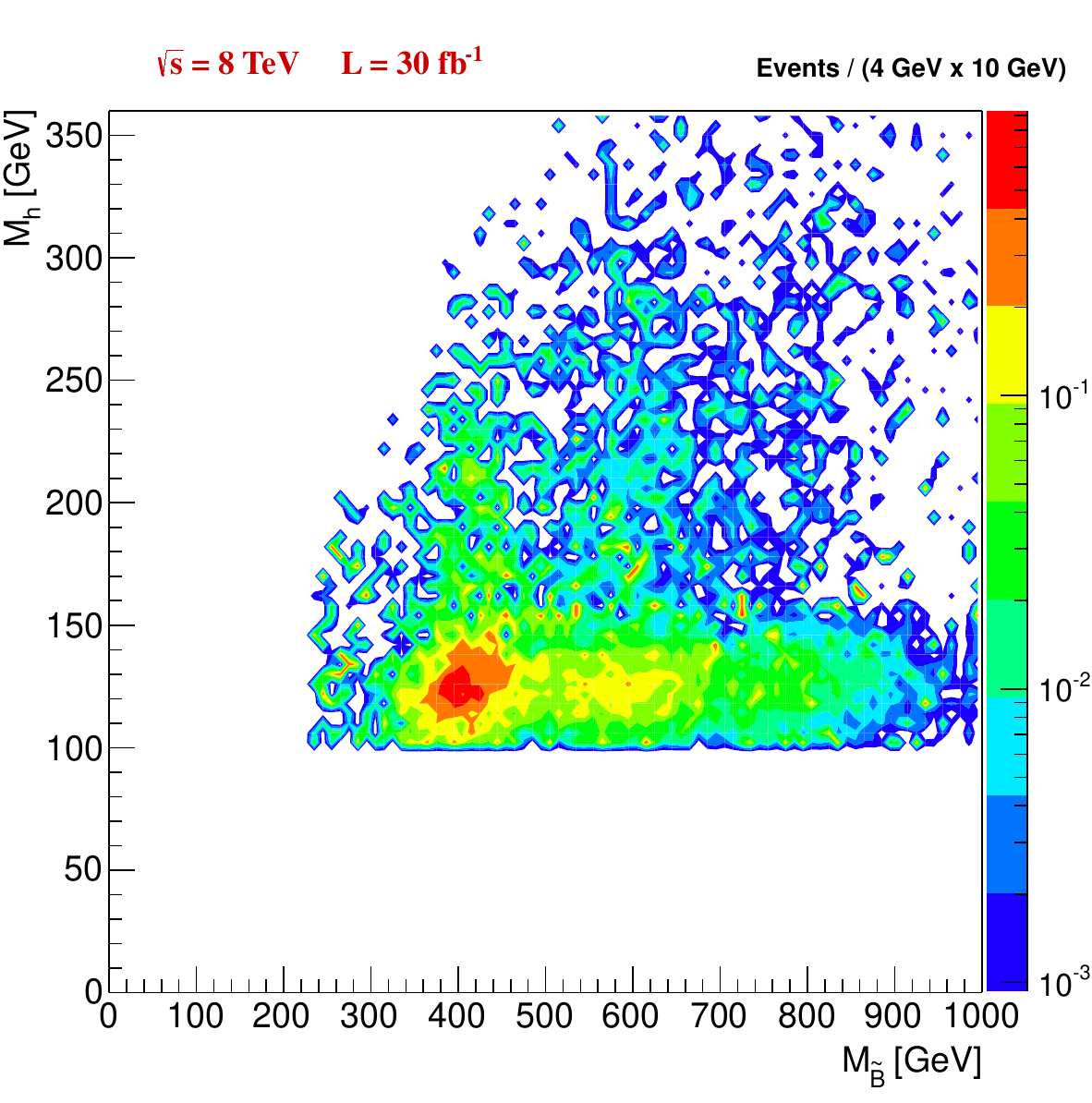}
\caption[]{
\label{fig:MhMBs}
\small
Contour plots of the invariant mass of the objects selected (with the procedure explained in Sec. \ref{sec:reco}) as the Higgs decay products, $M_h$, versus the invariant mass 
of the selected $\tilde{B}$ decay products, $M_{\tilde{B}}$, for the total background plus the signal at $m_h=125$ GeV and at different $\tilde{B}$ mass values ($400$ GeV, $600$ GeV and $800$ GeV), at $\sqrt{s}=8$ TeV. Left plot: signal and background events after the optimized cuts of eq. (\ref{eq:optimized});
Right plot: signal and background events after the optimized cuts of eq. (\ref{eq:optimized}) plus the cuts $M_{\tilde{B}}>230$ GeV and $M_h>100$ GeV.
}
\end{figure}

The cross sections for the signal and the main backgrounds after the optimized cuts of eq. (\ref{eq:optimized}) are reported in the column of Table \ref{tab:cutopt} labelled as \textsf{opt}. The left plot in Fig. \ref{fig:MhMBs} shows, after the optimized cuts, the contour plot of the invariant mass of the objects selected (with the procedure explained in Sec. \ref{sec:reco}) as the Higgs decay products, $M_h$, versus the invariant mass 
of the selected $\tilde{B}$ decay products, $M_{\tilde{B}}$, for the total background plus the signal at different $\tilde{B}$ mass values, at $\sqrt{s}=8$ TeV.
After the optimized cuts the background is substantially reduced. The residual background is mainly formed by $WWbb+jets$ events. We see, indeed, that the most part of the events is distributed on a region around $(m_t, m_W)$ in the contour plot $(M_{\tilde{B}},M_h)$ of Fig. \ref{fig:MhMBs} (left plot). The reconstruction we do tend to select, in the case of the $WWbb+jets$ background (which is made for the most part by $t\bar{t}+jets$ events) a top instead of a $\tilde{B}$ and the $W$ coming from this top instead of the Higgs. We see from the right plot on Fig. \ref{fig:MhMBs} that, after cutting away the $(m_t, m_W)$ region (we impose $M_{\tilde{B}}>230$ GeV and $M_{h}>100$ GeV) the background is strongly reduced and we can clearly distinguish the excess of events in correspondence of the heavy-bottom and of the Higgs resonances.
The cross sections for the signal and the background after the optimized cuts plus the cuts $M_{\tilde{B}}>230$ GeV and $M_{h}>100$ GeV are shown in the \textsf{opt+} column of Table \ref{tab:cutopt}.\\ 
Once we can recognize the heavy-bottom and the Higgs resonances we can also refine the analysis by imposing a cut on $M_{\tilde{B}}$ and on $M_h$. We require $M_h$ to be comprised in the region $[100$ GeV, $150$ GeV] and $M_{\tilde{B}}$ to be in a region of $\pm \Gamma(\tilde{B})$ from the $\tilde{B}$ mass value; 
for $\tilde{B}$ masses above $1$ TeV ($0.8$ TeV) in the case $\sqrt{s}=14$ TeV ($\sqrt{s}=8$ TeV) we impose a cut on $M_{\tilde{B}}$ of $\pm 2\Gamma(\tilde{B})$ from its central value. Indeed, for higher $\tilde{B}$ masses the background is already strongly reduced by a milder restriction on $M_{\tilde{B}}$. We also refine the cut on $M_{tot}$ and on the $p_T$ of the hardest jet according to the values shown on Table \ref{tab:refined}.


%

%

%

%
\begin{table}
\begin{center}
{\small
\begin{tabular}{|c|cc|cc|}
\multicolumn{1}{c}{} & \multicolumn{2}{c}{\textsf{LHC 8 TeV}} &   \multicolumn{2}{c}{\textsf{LHC 14 TeV}}\\[0.1cm]
\hline
 & \multicolumn{1}{c}{\textsf{opt}} & \multicolumn{1}{c|}{\textsf{opt+}} & \multicolumn{1}{c}{\textsf{opt}} & \multicolumn{1}{c|}{\textsf{opt+}} \\[0.1cm]
 & & & & \\[-0.3cm]
Signal $m_{\tilde{B}}=0.4\,$TeV & 1.7 & 1.5 & 14 & 13 \\[0.25cm]
 Signal $m_{\tilde{B}}=0.6\,$TeV & 0.56 & 0.52 & 5.1 & 4.7 \\[0.25cm]
 Signal $m_{\tilde{B}}=0.8\,$TeV & 0.15 & 0.15 & 1.8 & 1.7\\[0.25cm]
 Signal $m_{\tilde{B}}=1.0\,$TeV & & & 0.67 & 0.64 \\[0.25cm]
 Signal $m_{\tilde{B}}=1.2\,$TeV & & & 0.28 & 0.27 \\[0.35cm]
  $WWbb+jets$            & 5.2 & 1.1  & 58 & 11 \\[0.25cm]
  $Wbb+jets$                &  0.44  & 0.24 & 7.8 & 2.9 \\[0.25cm]
   $W+jets$              &  0.045 &  0.025 & 0.46 & 0.21\\[0.35cm]
  Total                  & & & &\\
  background       & 5.7 &  1.4 & 66 & 14   \\[0.15cm]
\hline
\end{tabular}
}
\caption{
\label{tab:cutopt}
\small 
Cross sections, in fb, for the signal (with $\lambda=3$) and the main backgrounds after the optimized cuts of eq. \ref{eq:optimized} (\textsf{opt}) and after the further cuts: $M_{\tilde{B}}>230$ GeV and $M_{h}>100$ GeV (\textsf{opt+}), for $\sqrt{s} = 8\,$TeV and $\sqrt{s} = 14\,$TeV.
}
\end{center}
\end{table}

%
%

%
\begin{table}
\begin{center}
{\small
\begin{tabular}{|c|c|c|c|c|c|}
\hline
&&&&&\\
  $m_{\tilde{B}}$ & 0.4 & 0.6 & 0.8  & 1.0 & 1.2 \\[0.15cm]
  \hline 
  &&&&& \\[0.05cm]
 $M_{tot}>$ & 1.2 & 1.3 & 1.4 & 1.7 & 1.8 \\ [0.17cm]
 $p_T$ $j(1)>$ & 0.17 & 0.24 & 0.30 & 0.42 & 0.46 \\ [0.17cm]
 $M_h \in$ & [0.10, 0.15]& [0.10, 0.15]& [0.10, 0.15]& [0.10, 0.15]& [0.10, 0.15]\\[0.17cm]
 $M_{\tilde{B}} \in$ & 0.4 $\pm\Gamma(\tilde{B})$& 0.6 $\pm\Gamma(\tilde{B})$ & 0.8 $\pm\Gamma(\tilde{B})$ (\textsf{14 TeV LHC}) & 1.0 $\pm 2\Gamma(\tilde{B})$& 1.2 $\pm 2\Gamma(\tilde{B})$\\
 & &  & 0.8 $\pm 2\Gamma(\tilde{B})$ (\textsf{8 TeV LHC}) & & \\
 &&&&&\\
 \hline
\end{tabular}
}
\caption{
\label{tab:refined}
\small 
Refined cuts, in TeV, for different $\tilde{B}$ mass values (first row), in TeV.
}
\end{center}
\end{table}

The final cross sections for the signal and the main backgrounds, after imposing the optimized cuts of eq.(\ref{eq:optimized}) plus the refined cuts of Table \ref{tab:refined}, are reported in Table \ref{tab:cutflow8TeV2} and \ref{tab:cutflow14TeV2} respectively for $\sqrt{s}=8\,$TeV and $\sqrt{s}=14\,$TeV. 
The values of the corresponding discovery luminosity are shown in Table~\ref{tab:Ldisc}.

For the background, we indicate in parenthesis the one-sigma statistical error on the cross section; 
for the signal, the statistical error is negligible, compared to that of the background, and we do not report it.
Statistical errors on the cross sections are computed by assuming a Poisson distribution for the number of events that pass the cuts.\footnote{We calculate the cross section after the application of a cut as
\[
 \sigma= \frac{n}{L}\ ,
\]
where $n$ is the number of simulated events that pass the cut and $L$ is the integrated luminosity reached in the simulation. Given the observed number of events, $n$, the true value of the number of events passing the cut, $\lambda$, follows a Poisson distribution:
\[
 f(\lambda|n)=\frac{\lambda e^{-\lambda}}{n!}\ .
\]
the variance associated with $\lambda$ is $Var[\lambda]=n+1$, we thus associate to the cross section a variance:
\[
 Var[\sigma]=\frac{n+1}{L^2} \ .
\]
\noindent
When we sum over different cross section values, the error is summed in quadrature.}
In order to obtain a conservative estimate of the discovery luminosity, we consider the central value plus one-sigma as the value of the background cross section.

\begin{table}
\begin{center}
{\small
\begin{tabular}{|c|c|c|c|}
\hline 
 & & &  \\[-0.35cm]
\multicolumn{1}{|l|}{\textsf{LHC  $\mathsf{8\,}$TeV}}& \multicolumn{1}{l|}{$m_{\tilde{B}} = 0.4\,$TeV} & \multicolumn{1}{l|}{$m_{\tilde{B}} = 0.6\,$TeV} & \multicolumn{1}{l|}{$m_{\tilde{B}} = 0.8\,$TeV} \\[0.1cm]
\hline
& & &   \\[-0.3cm]

Signal                 & 0.788 & 0.234 & 0.073 \\[0.25cm]
$WWbb+jets$     & 0.10(2)  & 0.02(1) & 0.02(1)\\[0.15cm]
$Wbb+jets$         & 0.017(4) & 0.011(3) & 0.013(4)  \\[0.15cm] 
$W+jets$             & 0.0015(3) & 0.0013(3) & 0.0014(3)\\[0.15cm]  

Total & & &  \\
background      & 0.12(2) & 0.04(1) & 0.04(1)\\[0.1cm]
\hline
\end{tabular}
\caption{
\label{tab:cutflow8TeV2}
\small 
Cross sections, in fb, at $\sqrt{s}=8\,$TeV for the signal (with $\lambda=3$) and the main backgrounds after imposing the optimized cuts of eq.(\ref{eq:optimized}) plus the refined cuts of Table \ref{tab:refined}. The statistical errors on the cross sections have been computed as explained in the text.
}}
\end{center}
\end{table}
\begin{table}
\begin{center}
{\small
\begin{tabular}{|c|c|c|c|c|c|}
\hline 
 & & & & &  \\[-0.35cm]
\multicolumn{1}{|l|}{\textsf{LHC  $\mathsf{14\,}$TeV}}& \multicolumn{1}{l|}{$m_{\tilde{B}} = 0.4\,$TeV} & \multicolumn{1}{l|}{$m_{\tilde{B}} = 0.6\,$TeV} & \multicolumn{1}{l|}{$m_{\tilde{B}} = 0.8\,$TeV} & \multicolumn{1}{l|}{$m_{\tilde{B}} = 1.0\,$TeV} & \multicolumn{1}{l|}{$m_{\tilde{B}} = 1.2\,$TeV}\\[0.1cm]
\hline
& & &  & &  \\[-0.3cm]

Signal                & 6.65 & 2.15 & 0.777 & 0.248 & 0.110 \\[0.25cm]
$WWbb+jets$    & 0.9(1) & 0.16(5) & 0.16(5) & 0.04(3) & 0.04(3) \\[0.15cm]  
$Wbb+jets$      &   0.16(2) & 0.14(2) & 0.10(2) & 0.06(1) &  0.020(8) \\[0.15cm] 
$W+jets$           &  0.012(2) & 0.011(2) & 0.007(2) & 0.002(1) & 0.0011(9) \\[0.15cm] 

Total & & & & &  \\
background     & 1.0(1) & 0.31(5) & 0.27(5) & 0.09(3) & 0.06(3) \\[0.1cm]
\hline
\end{tabular}
\caption{
\label{tab:cutflow14TeV2}
\small 
Cross sections, in fb, at $\sqrt{s}=14\,$TeV for the signal (with $\lambda=3$) and the main backgrounds after imposing the optimized cuts of eq.(\ref{eq:optimized}) plus the refined cuts of Table \ref{tab:refined}. The statistical errors on the cross sections have been computed as explained in the text.
}}
\end{center}
\end{table}

%

\begin{table}[h]
\begin{center}
{\small
\vspace{0.5cm}
\begin{tabular}{r|ccccc}
\multirow{2}{*}{\textsf{LHC $\mathsf{\sqrt{s} = 8\,}$TeV}} & \multicolumn{3}{c}{$m_{\tilde{B}} [\text{TeV}]$} & \\[0.15cm]
  & 0.4 & 0.6 & 0.8 & \\[0.1cm]
\cline{1-4}
& & & & \\[-0.3cm]
$L_{disc} \,[\text{fb}^{-1}]$ & 14 & 50 & 340 &
\\
\multicolumn{4}{c}{}
\\[0.7cm]
\multirow{2}{*}{\textsf{LHC $\mathsf{\sqrt{s} = 14\,}$TeV}} & \multicolumn{5}{c}{$m_{\tilde{B}} [\text{TeV}]$} \\[0.15cm]
 & 0.4 & 0.6 & 0.8 & 1.0 & 1.2 \\[0.1cm]
\hline
& & & & &\\[-0.3cm]
$L_{disc} \,[\text{fb}^{-1}]$ & 1.5 & 4.8 & 23 & 78 & 260
\end{tabular}
\vspace{0.5cm}
\caption{
\label{tab:Ldisc}
\small 
Value of the integrated luminosity required for a $5\sigma$ discovery after the optimized cuts of eq.(\ref{eq:optimized}) plus the refined cuts of Table \ref{tab:refined}, 
for $\sqrt{s}=8\,$TeV (upper panel) and $\sqrt{s}=14\,$TeV (lower panel).}}
\end{center}
\end{table}

\subsection{Discovery reach on the parameter space}

All the numbers shown in Tables~\ref{tab:cutflow8TeV2}, \ref{tab:cutflow14TeV2} and~\ref{tab:Ldisc} hold for a fixed coupling $\lambda=3$. 
It is very interesting to study the dependence of our results on the coupling $\lambda$; this, indeed, could give us 
an estimate of the LHC sensitivity to measure the Higgs (and electro-weak bosons) coupling to the heavy fermion, for different masses of this latter, and, consequently, to obtain a hint on the value of the Yukawa coupling among composites ($Y_{*}$), depending on the top degree of compositeness ($s_1$). 
We can generalize our results to different $\lambda$ values, by simply considering that the production cross section scales with $\lambda^2$. $\tilde{B}$ BRs do not depend on $\lambda$ and, a part from the production cross section, the only residual dependence of our results on $\lambda$ is in the $\tilde{B}$ total decay width, which depends on $\lambda$ quadratically.
It is thus possible to estimate how the LHC discovery reach
varies with $\lambda$ by simply rescaling the numbers in Tables~\ref{tab:cutflow8TeV2}, 
\ref{tab:cutflow14TeV2} to take into account the change in the production cross section.  

\begin{figure}[tbp]
\begin{center}
\hspace*{0.8cm}
\includegraphics[width=0.99\textwidth,clip,angle=0]{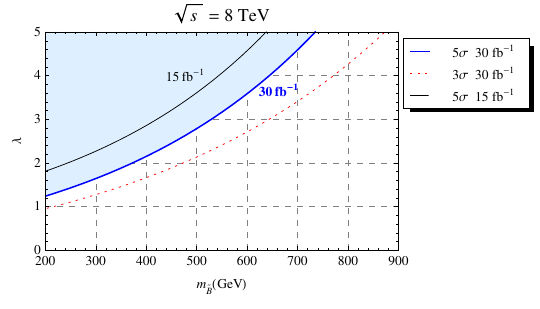}
\\[0.5cm]
\includegraphics[width=0.88\textwidth,clip,angle=0]{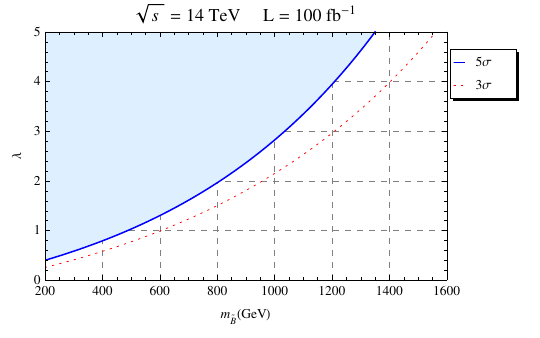}
\caption[]{
\label{fig:reach}
\small
LHC discovery reach in the plane $(m_{\tilde{B}},\lambda)$. The blue area shows the region where a discovery of the 
signal $pp\to (\tilde{B}\to (h\to bb)b)t+X$ is possible at $5\sigma$.
The $3\sigma$ reach is shown by the dotted red curve.
Upper plot: LHC at $\sqrt{s}=8\,$TeV with an integrated luminosity $L = 30\,\text{fb}^{-1}$; the $5\sigma$ reach at $L = 15\,\text{fb}^{-1}$ is also plotted and shown by the continuous black line; Lower plot:
 LHC at $\sqrt{s}=14\,$TeV with  $L = 100\,\text{fb}^{-1}$.
}
\end{center}
\end{figure}

The result is reported in Fig.~\ref{fig:reach}. The two plots show the 
region in the plane $(m_{\tilde{B}},\lambda)$ where a $5\sigma$ discovery is possible for the LHC at $\sqrt{s}=8\,$TeV with $L = 30\,\text{fb}^{-1}$ and $L = 15\,\text{fb}^{-1}$
(upper plot), and at $\sqrt{s}=14\,$TeV  with $L = 100\,\text{fb}^{-1}$ (lower plot).\\

We did not take into account the variation with $\lambda$ of the $\tilde{B}$ total decay width. In the region with $\lambda>3$ this latter is larger than the values considered in the analysis. In such region, however, the significance goes down to values below $5\sigma$ only in correspondence of the highest $\tilde{B}$ masses, for which we applied a quite mild cut, of $\pm 2 \Gamma(\tilde{B})$, on the invariant mass $M_{\tilde{B}}$. 
We thus believe that Fig. \ref{fig:reach} traces a conservative picture of the LHC reach.

\section{Discussion}\label{sec:conclusions}

Our results are summarized by Fig. \ref{fig:reach}. They show that, for a reference value $\lambda=3$ of the Higgs coupling to a heavy bottom $\tilde{B}$, the 8 TeV LHC with $30$ fb$^{-1}$ can discover a Higgs from a singly-produced heavy bottom if this latter has a mass $m_{\tilde{B}}\lesssim 530$ GeV (while an observation is possible for $m_{\tilde{B}}\lesssim 650$ GeV). 
These values for the $\tilde{B}$ mass are realistic, for example, in the case of a right-handed top with a large degree of compositeness. A possible composite nature of the Higgs could be thus revealed in the $(\tilde{B} \to (h\to bb)b)t+X$ channel already in the current phase of the LHC run. The LHC reach on the plane $(m_{\tilde{B}}, \lambda)$ is wide at $\sqrt{s}=14$ TeV. With $\lambda=3$ the LHC with $100$ fb$^{-1}$ can discover a Higgs from a heavy bottom for masses of this latter up to $\simeq 1040$ GeV. In the case of a custodian heavy bottom as light as $\simeq 500$ GeV, the 14 TeV LHC is sensitive to the measure of the $\lambda$ coupling in basically the full range $\lambda>1$ predicted by the theory.\\
The single production of a heavy fermion is thus a very promising channel to observe the Higgs and to test its possible composite nature.  The $\tilde{B}t+X$, in particular, proves to be an interesting channel also for the discovery of the heavy bottom. Since $BR(\tilde{B}\to hb)\simeq BR(\tilde{B}\to Zb)$, results similar to those obtained in this work are expected from the $(\tilde{B} \to (Z\to hadrons)b)t+X$ channel if one adopts the strategy outlined here, with a variation in the cut on the Higgs ($Z$) invariant mass.\\ 
In this work we have studied the single production of a heavy bottom; further contributions to the composite Higgs production are expected from other heavy fermions.  
In Table \ref{tab:channels} we finally list all the channels where heavy fermions can mediate the production of a composite Higgs in the TS10; similar scenarios are expected in different composite Higgs models with custodial symmetry. For each final state we indicate the type of mediating heavy fermion and the condition when this latter could be light. Comparative analyses of these channels can shed light on the theory and on the EWSB mechanism; The measure of the $\lambda$ couplings of the Higgs to different heavy fermions could give the possibility of extracting the value of the Yukawa coupling among composite states and the top degree of compositeness (see the expressions for the $\lambda$ couplings in eq. (\ref{eq:vertex}) of the Appendix \ref{TS10A}). \\






\begin{table}
\begin{center}
{\small
\begin{tabular}{|c|c|c|}
\hline 
 & &   \\[-0.35cm]
\multicolumn{1}{|c|}{\textsf{Final state}}& \multicolumn{1}{c|}{\textsf{Mediating heavy fermion}} & \multicolumn{1}{c|}{\textsf{light for}} \\[0.2cm]
\hline
&&\\[0.1cm]
$\mathbf{hbt+jets}$ \qquad \footnotesize{($W_L$ exchange)} & $\tilde{B}$ \qquad \footnotesize{$BR(\tilde{B}\to hb)\simeq 25\%$} & \footnotesize{composite $t_R$}    \\[0.1cm]
& $\tilde{T}$ \qquad \footnotesize{ $BR(\tilde{T}\to ht)\simeq 25\%$} & \\[0.1cm]
& $\tilde{T}^{'}$  \qquad \footnotesize{$BR(\tilde{T}^{'}\to ht)\simeq 25\%$} & \footnotesize{composite $t_R$} \\[0.1cm]
\hline
&&\\[0.1cm]
$\mathbf{ht\bar{t}+jets}$ \qquad \footnotesize{($Z_L/h$ exchange)} & $T $  \qquad \footnotesize{$BR(T\to ht)\simeq 50\%$} &    \\[0.1cm]
 & $T_{2/3}$  \qquad \footnotesize{$BR(T_{2/3}\to ht)\simeq 50\%$} &   \footnotesize{composite $t_L$} \\[0.1cm]
 & $\tilde{T}$ \qquad \footnotesize{ $BR(\tilde{T}\to ht)\simeq 25\%$} & \\[0.1cm]
& $\tilde{T}^{'}$  \qquad \footnotesize{$BR(\tilde{T}^{'}\to ht)\simeq 25\%$} & \footnotesize{composite $t_R$} \\[0.1cm]
\hline
&&\\[0.1cm]
$\mathbf{hb\bar{b}+jets}$ \qquad \footnotesize{($Z_L/h$ exchange)} & $\tilde{B}$ \qquad \footnotesize{ $BR(\tilde{B}\to hb)\simeq 25\%$} & \footnotesize{composite $t_R$ }   \\[0.1cm]
& $\tilde{B}^{'}$  \qquad \footnotesize{$BR(\tilde{B}^{'}\to hb)\simeq 50\%$} & \footnotesize{composite $t_R$} \\[0.1cm]
\hline
\end{tabular}
\caption{
\label{tab:channels}
\small 
Pattern of the contribution to the Higgs production from different singly-produced heavy fermions in the TS10. 
We indicate the final state, the type of mediating heavy fermion and the condition when this latter could be light. 
}}
\end{center}
\end{table}

\section*{Acknowledgments}
I thank Roberto Contino for discussions and comments.
This work was supported in part by DOE under contract number DE-FG02-01ER41155.


\appendix
\section*{Appendix}

\section{TS10}
\label{TS10A}

Fermions rotate from the elementary/composite basis to the physical light(SM)/heavy basis as:
\begin{align}
\begin{split}\label{rotation_LL_TS}
& \tan\varphi_{L}=\frac{\Delta_{L}}{M_{Q*}}\equiv \frac{s_{1}}{c_{1}}\\  
& \left\{\begin{array}{l}
	t_{L}=c_{1}t^{el}_{L}-s_{1}T^{com}_{L}\\
	T_{L}=s_{1}t^{el}_{L}+c_{1}T^{com}_{L}
\end{array} \right.  \ \ 
 \left\{\begin{array}{l}
	b_{L}=c_{1}b^{el}_{L}-s_{1}B^{com}_{L}\\
	B_{L}=s_{1}b^{el}_{L}+c_{1}B^{com}_{L}  
\end{array} \right.
\end{split}
\end{align}

\begin{align}
\begin{split}\label{rotation_RR_TS}
& \tan\varphi_{R}=\frac{\Delta_{R1}}{M_{\tilde{Q}*}}\ \ s_{R}\equiv\sin\varphi_{R} \ \ c_{R}\equiv\cos\varphi_{R} \\
& \tan\varphi_{bR}=\frac{\Delta_{R2}}{M_{\tilde{Q}*}}\ \ s_{bR}\equiv\sin\varphi_{bR} \ \ c_{bR}\equiv\cos\varphi_{bR} \\
& \left\{\begin{array}{l}
	t_{R}=c_{R}t^{el}_{R}-s_{R}\tilde{T}^{com}_{R}\\
	\tilde{T}_{R}=s_{R}t^{el}_{R}+c_{R}\tilde{T}^{com}_{R} 
\end{array}  \right.  \ \ 
 \left\{\begin{array}{l}
	b_{R}=c_{bR}b^{el}_{R}-s_{bR}\tilde{B}^{com}_{R} \\ 
	\tilde{B}_{R}=s_{bR}b^{el}_{R}+c_{bR}\tilde{B}^{com}_{R}
\end{array} \right. 
\end{split}
\end{align}

\noindent
Physical heavy fermion masses are related to the bare ones as:
\begin{align}
\left\{\begin{array}{l}
m_{\tilde{T}}=\sqrt{M^{2}_{\tilde{Q}*}+\Delta^{2}_{R1}}=\frac{M_{\tilde{Q}*}}{c_{R}}\\
m_{\tilde{B}}=\sqrt{M^{2}_{\tilde{Q}*}+\Delta^{2}_{R2}}=\frac{M_{\tilde{Q}*}}{c_{bR}}\\
m_{\tilde{T}5/3}=m_{\tilde{T}'5/3}=m_{\tilde{T}'}=m_{\tilde{B}'}=M_{\tilde{Q}*}\\
m_{T}=m_B=\sqrt{M^{2}_{Q*}+\Delta^{2}_{L}}=\frac{M_{Q*}}{c_{1}} \\
m_{T2/3}=m_{T5/3}=M_{Q*} 
\end{array} \right.
\end{align}\\

\noindent
After field rotation to the mass eigenstate basis (eq.s (\ref{rotation_LL_TS}) and (\ref{rotation_RR_TS})), and before EWSB, the final Yukawa Lagrangian reads:



\begin{align}
\begin{split}\label{eq.Lagrange2_ts10}	
\mathcal{L}^{YUK}= & Y_{*}c_{1}c_{R}\frac{1}{\sqrt{2}}\left(\bar{T}_{L}\phi^{\dag}_{0}\tilde{T}_{R}-\bar{B}_{L}\phi^{-}\tilde{T}_{R}\right)-Y_{*}c_{R}\frac{1}{\sqrt{2}}\left(\bar{T}_{2/3L}\phi_{0}\tilde{T}_{R}+\bar{T}_{5/3L}\phi^{+}\tilde{T}_{R}\right)\\ 
& - Y_{*}s_{1}c_{R}\frac{1}{\sqrt{2}}\left(\bar{t}_{L}\phi^{\dag}_{0}\tilde{T}_{R}-\bar{b}_{L}\phi^{-}\tilde{T}_{R}\right)+Y_{*}s_{1}s_{R}\frac{1}{\sqrt{2}}\left(\bar{t}_{L}\phi^{\dag}_{0}t_{R}-\bar{b}_{L}\phi^{-}t_{R}\right)\\ 
& +Y_{*}s_{R}\frac{1}{\sqrt{2}}\left(\bar{T}_{2/3L}\phi_{0}t_{R}+\bar{T}_{5/3L}\phi^{+}t_{R}\right)
 -Y_{*}c_{1}s_{R}\frac{1}{\sqrt{2}}\left(\bar{T}_{L}\phi^{\dag}_{0}t_{R}-\bar{B}_{L}\phi^{-}t_{R}\right)\\ 
& +Y_{*}\frac{1}{\sqrt{2}}\left(\bar{T}_{R}\phi^{\dag}_{0}\tilde{T}_{L}-\bar{B}_{R}\phi^{-}\tilde{T}_{L}\right)-Y_{*}\frac{1}{\sqrt{2}}\left(\bar{T}_{2/3R}\phi_{0}\tilde{T}_{L}+\bar{T}_{5/3R}\phi^{+}\tilde{T}_{L}\right)\\ 
& +Y_{*}\left(\bar{T}_{5/3L}\phi^{\dag}_{0}\tilde{T}_{5/3R}-\bar{T}_{2/3L}\phi^{-}\tilde{T}_{5/3R}\right)+Y_{*}\left(\bar{T}_{5/3R}\phi^{\dag}_{0}\tilde{T}_{5/3L}- \bar{T}_{2/3R}\phi^{-}\tilde{T}_{5/3L}\right)\\
&-Y_{*}s_{1}c_{bR}\left(\bar{b}_{L}\phi_{0}\tilde{B}_{R}+\bar{t}_{L}\phi^{+}\tilde{B}_{R}\right)+Y_{*}s_{1}s_{bR}\left(\bar{b}_{L}\phi_{0}b_{R}+\bar{t}_{L}\phi^{+}b_{R}\right)\\ 
&-Y_{*}c_{1}s_{bR}\left(\bar{B}_{L}\phi_{0}b_{R}+\bar{T}_{L}\phi^{+}b_{R}\right)+Y_{*}c_{1}c_{bR}\left(\bar{B}_{L}\phi_{0}\tilde{B}_{R}+\bar{T}_{L}\phi^{+}\tilde{B}_{R}\right)\\
&+Y_{*}\left(\bar{B}_{R}\phi_{0}\tilde{B}_{L}+\bar{T}_{R}\phi^{+}\tilde{B}_{L}\right)+ Y_{*}\left(\bar{B}_{R}\phi^{\dag}_{0}\tilde{B}'_{L}+Y_{*}\bar{T}_{2/3R}\phi^{+}\tilde{B}'_{L}\right)\\
& Y_{*}\frac{1}{\sqrt{2}}\left(\bar{T}_{R}\phi^{\dag}_{0}\tilde{T}'_{L}+\bar{B}_{R}\phi^{-}\tilde{T}'_{L}\right)-Y_{*}\frac{1}{\sqrt{2}}\left(\bar{T}_{2/3R}\phi_{0}\tilde{T}'_{L}-\bar{T}_{5/3R}\phi^{+}\tilde{T}'_{L}\right)\\
& + Y_{*}c_{1}\frac{1}{\sqrt{2}}\left(\bar{T}_{L}\phi^{\dag}_{0}\tilde{T}'_{R}+\bar{B}_{L}\phi^{-}\tilde{T}'_{R}\right)-Y_{*}\frac{1}{\sqrt{2}}\left(\bar{T}_{2/3L}\phi^{\dag}_0\tilde{T}'_{R}-\bar{T}_{5/3L}\phi^{+}\tilde{T}'_{R}\right)\\
&- Y_{*}s_{1}\frac{1}{\sqrt{2}}\left(\bar{t}_{L}\phi^{\dag}_{0}\tilde{T}'_{R}+\bar{b}_{L}\phi^{-}\tilde{T}'_{R}\right)+Y_{*}\left(\bar{T}_{5/3R}\phi_{0}\tilde{T}'_{5/3L}-\bar{T}_{R}\phi^{-}\tilde{T}'_{5/3L}\right)\\
&+Y_{*}c_{1}\left(\bar{B}_{L}\phi^{\dag}_{0}\tilde{B}'_{R}-\bar{T}_{L}\phi^{-}\tilde{T}'_{5/3R}\right)-Y_{*}s_{1}\left(\bar{b}_{L}\phi^{\dag}_{0}\tilde{B}'_{R}-\bar{t}_{L}\phi^{-}\tilde{T}'_{5/3R}\right)\\
&+Y_{*}\bar{T}_{2/3L}\phi^{+}\tilde{B}'_{R}+Y_* \bar{T}_{5/3L}\phi_{0}\tilde{T}'_{5/3R} + h.c.
\end{split}
\end{align}
\noindent

\subsection{Heavy fermion decays}\label{BRsHf}

Heavy fermions are essentially composite states, therefore they couple strongly to composite modes.
Heavy fermions ($\chi$) decay thus into longitudinally polarized bosons or into the Higgs (plus a SM fermion $\psi$);
 the widths for these decays are as follows:

\[
\Gamma\left(\chi\rightarrow W_L\psi\right)=\frac{\lambda^2_{W\chi}}{32 \pi}M_{\chi}
\left[\left( 1+\frac{m^2_{\psi}-M^2_W}{M^2_{\chi}}\right)\left( 1+\frac{m^2_{\psi}+2M^2_W}{M^2_{\chi}}\right)-4\frac{m^2_{\psi}}{M^2_{\chi}}\right] 
\]
\[
 \times\sqrt{1-2\frac{m^2_{\psi}+M^2_W}{M^2_{\chi}}+\frac{\left(m^2_{\psi}-M^2_W\right)^2 }{M^4_{\chi}}}
\]

\[
\Gamma\left(\chi\rightarrow Z_L\psi\right)=\frac{\lambda^2_{Z\chi}}{64 \pi}M_{\chi}
\left[\left( 1+\frac{m^2_{\psi}-M^2_Z}{M^2_{\chi}}\right)\left( 1+\frac{m^2_{\psi}+2M^2_Z}{M^2_{\chi}}\right)-4\frac{m^2_{\psi}}{M^2_{\chi}}\right] 
\]
\[
 \times\sqrt{1-2\frac{m^2_{\psi}+M^2_Z}{M^2_{\chi}}+\frac{\left(m^2_{\psi}-M^2_Z\right)^2 }{M^4_{\chi}}}
\]

\[
\Gamma\left(\chi\rightarrow h\psi\right)=\frac{\lambda^2_{h\chi}}{64 \pi}M_{\chi}
\left( 1+\frac{m^2_{\psi}}{M^2_{\chi}}-\frac{M^2_{h}}{M^2_{\chi}}\right)
\sqrt{\left(1-\frac{m^2_{\psi}}{M^2_{\chi}}+\frac{M^2_{h}}{M^2_{\chi}}\right)^2-4\frac{M^2_{h}}{M^2_{\chi}}}\ .
\] \\

\noindent
Vertices $\lambda_{W/Z/h\chi}$ are:

\begin{align}
\label{eq:vertex}
 \begin{split}
&\text{for the $T$ decays ($T\to Wb$, $T\to Zt$, $T\to ht$)}\\[0.2cm]
& \hspace*{2cm} \lambda_{WT}\simeq 0 \ , \qquad \lambda_{ZT}=\lambda_{hT}=Y_* c_1s_R \\[0.5cm]
& \text{for the $B$ decays ($B\to Wt$, $B\to Zb$, $B\to hb$)} \\[0.2cm]
& \hspace*{2cm}\lambda_{WB}= Y_* c_1s_R \ , \qquad  \lambda_{ZB}\simeq 0 \ , \qquad  \lambda_{hB}\simeq 0 \\[0.5cm]
& \text{for the $\tilde{T}$ decays ($\tilde{T}\to Wb$, $\tilde{T}\to Zt$, $\tilde{T}\to ht$) } \\[0.2cm] &  \hspace*{2cm}\lambda_{W\tilde{T}}=\lambda_{Z\tilde{T}}=\lambda_{h\tilde{T}}= Y_* s_1 c_R  \\[0.5cm]
& \text{for the $\tilde{B}$ decays ($\tilde{B}\to Wt$, $\tilde{B}\to Zb$, $\tilde{B}\to hb$) } \\[0.2cm]  & \hspace*{2cm}\lambda=\lambda_{W\tilde{B}}=\lambda_{Z\tilde{B}}=\lambda_{h\tilde{B}}=Y_* s_1 c_{bR}\simeq Y_* s_1 
 \end{split}
\end{align}

\begin{figure}[tbp]
\begin{center}
\includegraphics[width=0.49\textwidth,clip,angle=0]{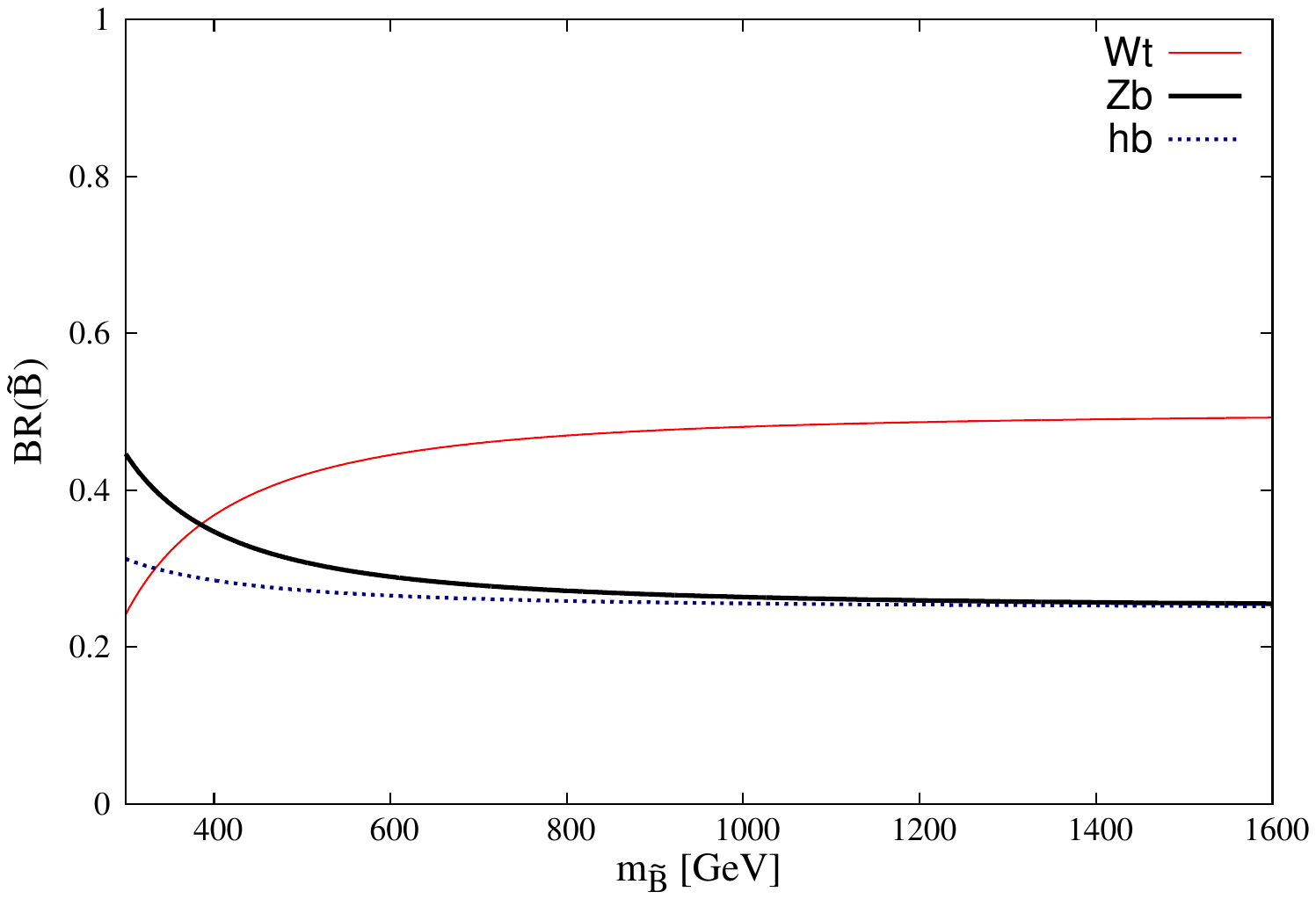}
\includegraphics[width=0.49\textwidth,clip,angle=0]{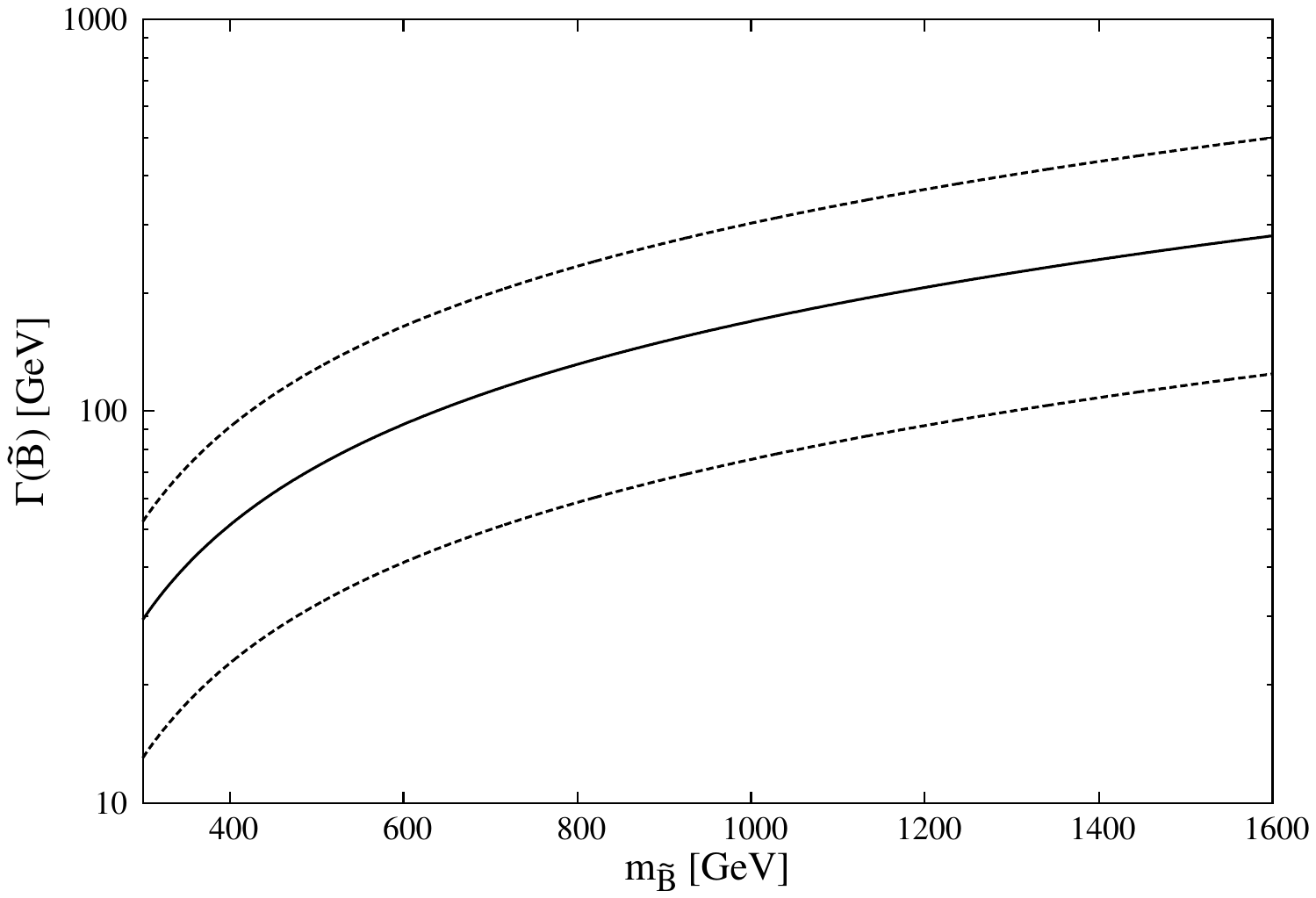}
\caption[]{
\label{fig:Bs-decay}
\small
BRs (Left Plot) and total decay width (Right Plot) of $\tilde{B}$. 
The width depends quadratically on $\lambda$; 
the continuous line in the Right Plot refers to $\lambda=3$, the dotted lines define a range of variation $2<\lambda<4$ of the total decay width. 
}
\end{center}
\end{figure}

\noindent
Fig. \ref{fig:Bs-decay} shows the BRs (Left Plot) and the total decay width (Right Plot) of $\tilde{B}$. 
The width depends quadratically on $\lambda$; 
the continuous line refers to $\lambda=3$, the dotted lines define a range of variation $2<\lambda<4$ of the total decay width.

%



\newpage

\begin{thebibliography}{99}


\bibitem{lhc-higgs} CMS Collaboration, 
 arXiv:1202.1488 [hep-ex]; ATLAS Collaboration, 
  arXiv:1202.1408 [hep-ex].

\bibitem{Georgi_Kaplan} D. B. Kaplan and H. Georgi, Phys. Lett. B {\bf 136}, 183 (1984).


\bibitem{Contino:2003ve}
  R.~Contino, Y.~Nomura, A.~Pomarol,
  Nucl.\ Phys.\ B  {\bf 671 } (2003)  148-174.
  [hep-ph/0306259].

\bibitem{Agashe:2004rs}
  K.~Agashe, R.~Contino, A.~Pomarol,
  Nucl.\ Phys.\ B  {\bf 719 } (2005)  165-187.
  [hep-ph/0412089].



\bibitem{Mrazek:2009yu}
  J.~Mrazek, A.~Wulzer,
  Phys.\ Rev.\  D {\bf 81 } (2010)  075006.
  [arXiv:0909.3977 [hep-ph]].

\bibitem {Servant} R. Contino and G. Servant, JHEP {\bf 0806} (2008) 026, arxiv:0801.1679 [hep-ph].  

\bibitem{Willenbrock}
S. S. D. Willenbrock, D. A. Dicus., Phys. Rev. D {\bf 34}:155, 1986.



\bibitem{Contino:2006nn}
  R.~Contino, T.~Kramer, M.~Son and R.~Sundrum,
  JHEP {\bf 0705}, 074 (2007)
  [arXiv:hep-ph/0612180].

\bibitem{contino-rattazzi} R. Contino, C. Grojean, M. Moretti, F. Piccinini and R. Rattazzi,  JHEP {\bf 1005} (2010) 089, arXiv:1002.1011 [hep-ph]

\bibitem{higgs-bound}
A. Azatov, R. Contino, and J. Galloway arXiv:1202.3415 [hep-ph];  J. Espinosa, C. Grojean, and M. Muhlleitner arXiv:1202.1286 [hep-ph]; D. Carmi, A. Falkowski, E. Kuflik and T. Volansky arXiv:1202.3144 [hep-ph];  J. Espinosa, C. Grojean, M. Muhlleitner and M. Trott arXiv:1202.3697 [hep-ph].


\bibitem{lhcNPG} G. Brooijmans, B. Gripaios, F. Moortgat,  J. Santiago, P. Skands \textit{et al.}, Les Houches 2011: Physics at TeV Colliders New Physics Working Group Report, arXiv:1203.1488 [hep-ph].
\bibitem{Azatov2012} A. Azatov \textit{et al.}, arXiv:1204.0455 [hep-ph].
\bibitem{Santiago2012} A. Carmona, M. Chala, J. Santiago, arXiv:1205.2378 [hep-ph].
\bibitem{Bini} C. Bini, R. Contino, N. Vignaroli, JHEP { \bf 1201} (2012) 157, arxiv:1110.6058 [hep-ph]; N. Vignaroli, Nuovo Cim. C {\bf 034} (2012) 06 213-216, arXiv:1107.4558 [hep-ph];  \bibitem{Santiago} R.~Barcelo, A.~Carmona, M.~Chala, M.~Masip, J.~Santiago, Nucl. Phys. B {\bf 857} (2012) 172-184, 
  arXiv:1110.5914 [hep-ph].
  \bibitem{Kong} K. Kong, M. McCaskey, G. W. Wilson, arXiv:1112.3041 [hep-ph].




\bibitem{vignaroli_tesi}
N. Vignaroli, Ph. D. thesis, arxiv:1112.0218 [hep-ph].
\bibitem{vignaroli_bsGamma}
N. Vignaroli, arxiv:1204.0478 [hep-ph]. 

\bibitem{vignaroli_preparation} N. Vignaroli, arXiv:1207.0830 [hep-ph].


\bibitem{Little-Higgs} N. Arkani-Hamed, A. G. Cohen, E. Katz and A. E. Nelson, JHEP { \bf 0207}, 034 (2002)
[arXiv:hep-ph/0206021].

\bibitem{Little-Higgs-analyses} T. Han, H. E. Logan, B. McElrath, and L.-T. Wang, Phys. Rev. D { \bf 67} 095004, 2003, hep-ph/0301040.
M. Perelstein, M. E. Peskin, and A. Pierce. Phys. Rev. D { \bf 69} 075002, 2004, hep-ph/0310039.

\bibitem{Contino:2006qr}
  R.~Contino, L.~Da Rold and A.~Pomarol,
  Phys.\ Rev.\  D {\bf 75}, 055014 (2007)
  [arXiv:hep-ph/0612048].
  
\bibitem{Carena:2006bn}
  M.~S.~Carena, E.~Ponton, J.~Santiago and C.~E.~M.~Wagner,
  Nucl.\ Phys.\  B {\bf 759} (2006) 202
  [arXiv:hep-ph/0607106];


\bibitem{Agashe:2006at}
  K.~Agashe, R.~Contino, L.~Da Rold and A.~Pomarol,
  Phys.\ Lett.\  B {\bf 641} (2006) 62
  [arXiv:hep-ph/0605341].



\bibitem{Agashe:2003zs}
  K.~Agashe, A.~Delgado, M.~J.~May, R.~Sundrum,
  JHEP {\bf 0308 } (2003)  050.
  [hep-ph/0308036].
  
  
\bibitem{Kaplan:1991dc}
  D.~B.~Kaplan,
  Nucl.\ Phys.\  B {\bf 365} (1991) 259.


\bibitem{anastasiouRCM}
C. Anastasiou, E. Furlan, J. Santiago, Phys. Rev. D {\bf 79}, 075003 (2009) 
[arXiv:0901.2127]
\bibitem{Redi2011} M. Redi and A. Weiler, JHEP {\bf 1111} (2011) 108, arXiv:1106.6357 [hep-ph].
\bibitem{Atlas_mBs_limit} ATLAS Collaboration, CERN-PH-EP-2011-230, arxiv:1202.6540 [hep-ex].


\bibitem{MG-ME}
  J.~Alwall {\it et al.},
  JHEP {\bf 0709} (2007) 028
  [arXiv:0706.2334 [hep-ph]];
  F.~Maltoni and T.~Stelzer,
  JHEP {\bf 0302} (2003) 027
  [arXiv:hep-ph/0208156];
  T.~Stelzer and W.~F.~Long,
  Comput.\ Phys.\ Commun.\  {\bf 81} (1994) 357
  [arXiv:hep-ph/9401258].
  
  \bibitem{Feynrules}
  N. D. Christensen, C. Duhr, 
  Comput. Phys. Commun. {\bf 180} (2009) 1614-1641 [arXiv:0806.4194 [hep-ph]] 

\bibitem{Mangano:2002ea}
  M.~L.~Mangano, M.~Moretti, F.~Piccinini, R.~Pittau, A.~D.~Polosa,
  JHEP {\bf 0307 } (2003)  001.
  [hep-ph/0206293].



\bibitem{Aad:2008zzm}
  G.~Aad {\it et al.} [ ATLAS Collaboration ],
  JINST  3  (2008)  S08003.


\end{thebibliography}
\end{document}